\documentclass[12pt,titlepage]{article}
\usepackage{apalike}
\usepackage{psfig}

\newcommand{\half}{\mbox{\small$\frac{1}{2}$}}

\newcommand{\A}{\ensuremath{\mathcal{A}}}
\newcommand{\B}{\ensuremath{\mathcal{B}}}
\newcommand{\C}{\ensuremath{\mathcal{C}}}
\newcommand{\D}{\ensuremath{\mathcal{D}}}

\def\singlespacing{\baselineskip=12pt}

\begin{document}
\singlespacing


\begin{center}
\begin{large}
{\bf Analysis of Dynamic Brain Imaging Data\footnote{Submitted to {\it
The Biophysical Journal}}} \bigskip 

\end{large}
\bigskip

\bigskip

P.P.Mitra\footnote{Corresponding Author:\\
\hspace*{10mm}  Room 1D-268,\\
\hspace*{10mm}  Bell Laboratories, Lucent Technologies,\\
\hspace*{10mm}  700, Mountain Ave.,\\
\hspace*{10mm}  Murray Hill, NJ 07974}\  
        and B.Pesaran 
\smallskip

Bell Laboratories, Lucent Technologies\\
700, Mountain Ave.\\
Murray Hill, NJ 07974 
\end{center}

\begin{abstract}

Modern imaging techniques for probing brain function, including
functional Magnetic Resonance Imaging, intrinsic and extrinsic
contrast optical imaging, and magnetoencephalography, generate large
data sets with complex content.  In this paper we develop appropriate
techniques of analysis and visualization of such imaging data, in
order to separate the signal from the noise, as well as to
characterize the signal.  The techniques developed fall into the
general category of multivariate time series analysis, and in
particular we extensively use the multitaper framework of spectral
analysis.  We develop specific protocols for the analysis of fMRI,
optical imaging and MEG data, and illustrate the techniques by
applications to real data sets generated by these imaging modalities.
In general, the analysis protocols involve two distinct stages:
`noise' characterization and suppression, and `signal'
characterization and visualization.  An important general conclusion
of our study is the utility of a frequency-based representation, with
short, moving analysis windows to account for non-stationarity in the
data.  Of particular note are (a) the development of a decomposition
technique (`space-frequency singular value decomposition') that is
shown to be a useful means of characterizing the image data, and 
(b) the development of an algorithm, based on multitaper methods, 
for the removal of approximately periodic physiological artifacts
arising from cardiac and respiratory sources. 

\bigskip
\bigskip
\bigskip
\bigskip
\bigskip
\bigskip
\bigskip

\textbf{Keywords}:  Spectral Analysis, functional magnetic resonance
imaging (fMRI), optical imaging, magnetoencephalography (MEG),
multivariate time series analysis, singular value decomposition.

\end{abstract}

\section{Introduction}

The  brain constitutes a complex dynamical  system with a large number
of  degrees of    freedom,  so that   multichannel  measurements  are
necessary  to  gain a detailed  understanding  of its  behavior. Such
multi-channel measurements, made available by current instrumentation,
include  multi-electrode   recordings,   optical  brain images   using
intrinsic  \cite{Blasdel86}, \cite{Grinvald86}
or  extrinsic \cite{Cohen73}
contrast agents, functional magnetic resonance
imaging (fMRI) \cite{Ogawa92},\cite{Kwong92}
and magnetoencephalography  (MEG) \cite{Hamalainen93}. Due to  improvements
in the capabilities  of the measuring apparatus,  as well as growth in
computational  power and storage capacity,  the data sets generated by
these experiments   are  increasingly  large  and  more complex.   The
analysis and  visualization of such  multichannel data is an important
piece of the  associated research program, and is  the subject of this
paper. 

There are several common  problems associated with the different types
of  multichannel  data enumerated  above.   Firstly, preprocessing  is
necessary   to   remove  nuisance    components,  arising from    both
instrumental and physiological sources, from the data. Secondly, 
an  appropriate representation of  the  data for purposes  of
analysis  and visualization is necessary. Thirdly, there  is the 
task of extracting
any underlying simplicities  from  the signal,  mostly in  absence  of
strong models for the dynamics of the relevant parts  of the brain. If
there are simple features that  are hidden  in  the complexity of  the
data, then the analytical methodology should be such as to reveal such
features efficiently. 
 
With the current exponential growth in computational power and storage
capacity, it is increasingly possible to  perform the above steps in a
semi-automated way, and even  in real time. In fact,  this is almost a
pre-requisite to the success  of multi-channel measurements, since the
large dimensionality of the  data sets effectively preclude exhaustive
manual inspection by the  human experimenter. An  additional challenge
is to  perform the above  steps as far as possible  in real time, thus
allowing  quick feedback into the  experiment.  The intimate interplay
between the basic  experimental apparatus and semi-automated  analysis
and visualization is schematically   illustrated in Fig.1.   Note that
even given the increases in storage capacity, it  is desirable to have
ways of  compressing    the  data while   retaining   the  appropriate
information, so as to prevent saturation of the available storage. 

Problems such as  the above  are clearly  not unique to  neuroscience.
Automated analysis plays an important  role in the emergent discipline
of computational molecular biology.  Despite  the current relevance of
these problems, the appropriate analytical and computational tools are
in an early  stage of development.   In addition, investigators in the
field are  sometimes  unaware   of   the  appropriate  modern   signal
processing tools.  Since
little is understood about the detailed workings of the brain,
a straightforward exploratory approach  using crude analysis protocols
is  usually favored.     However,  given   the   increasing 
availability  of
computational   resources, this  unnecessarily  limits  the  degree of
knowledge that can be gained  from the data, and at  worst can lead to
erroneous  conclusions, for   example  when  statistical   methods are
applied inappropriately   (cf.\cite{Cleveland}  p.177).  On the  other
hand, a superficial   application   of complex signal   processing  or
statistical  techniques, can  lead  to results  that are difficult  to
interpret. 

An aspect of the the data in question that cannot be emphasized enough
is the fact that the data constitute time series, mostly multivariate. 
While techniques for treating static high-dimensional data are widely
known and appreciated, both in multivariate statistics and in the field
of pattern recognition, the techniques for treating time series data
are less well developed, except in special cases. We focus particularly 
on the data as multivariate time series. In this paper, our concerns are
twofold: (a) characterization and   removal of the  typical artifacts,
and (b) characterization  of underlying structure  of the signal  left
after removal of artifacts.  The paper is organized into two parts. 

\begin{figure}
\centerline{\hbox{\psfig{figure=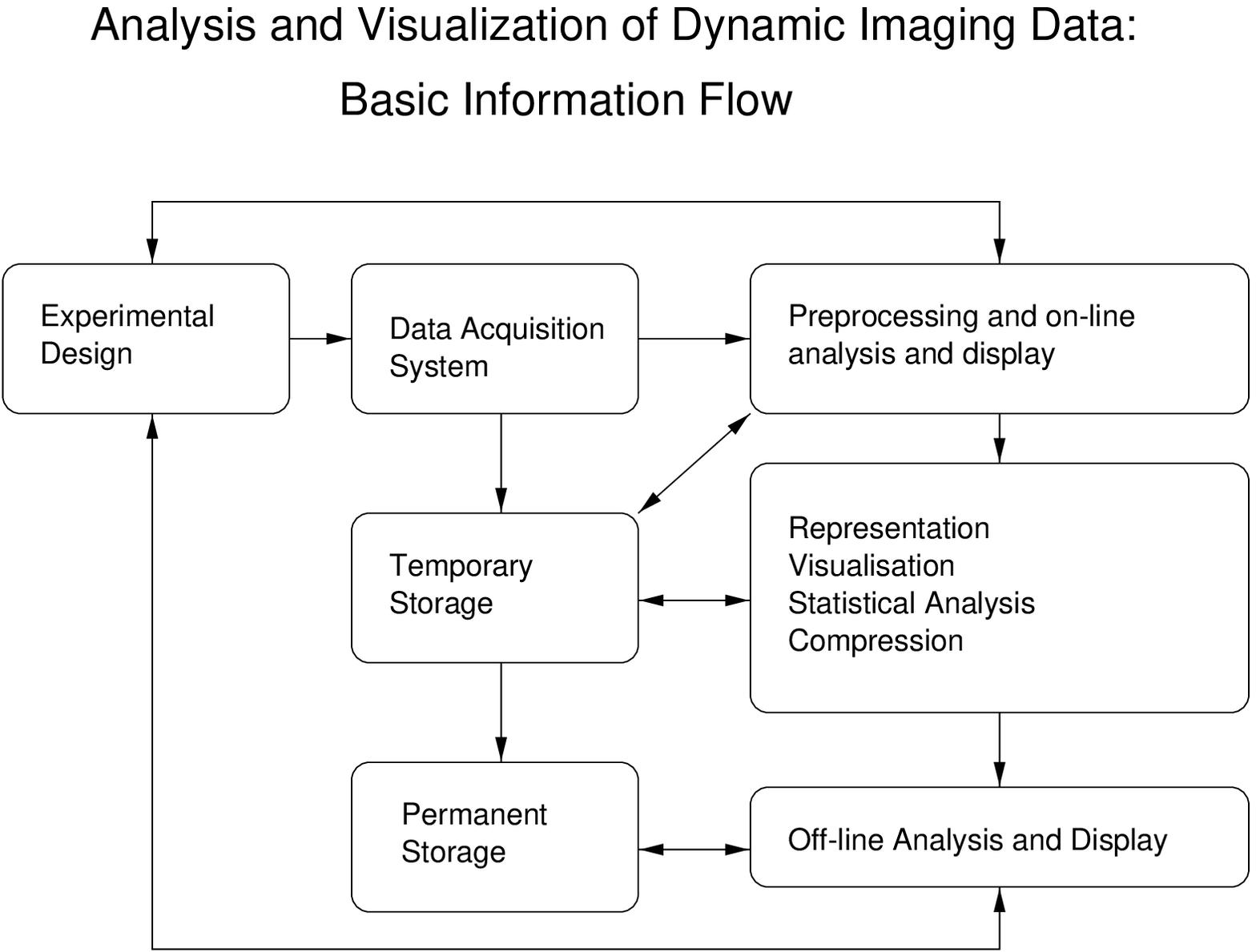,height=3in,width=4in}}}
\caption[abbrev]{
Schematic overview of data acquisition and analysis 
}
\end{figure}

In the first part we review some of the relevant analytical techniques
for multivariate time  series. In particular,  we provide a 
description  of multitaper spectral methods \cite{Thomson82},
\cite{Thomson91}, \cite{Percival93}.  This is a framework for performing
spectral analysis of univariate and multivariate  time series that has
particular advantages for the data  at hand. A central issue for the 
data we present is to be able to deal with very short data segments
and still obtain statistically well behaved estimators. Reasons for
this 
are that gathering long time series may be expensive (for example 
in fMRI), and that the presence of non-stationarity in the data makes
it preferable to use a short, moving analysis window. Multitaper 
methods are particularly powerful for performing spectral analysis 
of short data segments. In  the second part of  the paper, we treat  in
succession  dynamic  brain imaging data  gathered in  MEG, optical and
fMRI experiments. Based on the analysis of  actual data sets using the
techniques introduced  in the first  section, we discuss protocols for
analysis,   both to  remove artifactual    components  as well  as  to
determine the structure of the signal.

A  principal  motivation  for  the  current study    is the increasing
interest in  the   internal  dynamics  of  neural    systems.  In many
experimental  paradigms, past  or  present,  neural systems have  been
characterized  from the point of  view  of input-output relationships.
For example,  the quantity of interest is  often  a stimulus response,
and the  experiment  is performed by  repeating the  stimulus multiple
times, and the different trials are  regarded as forming a statistical
ensemble. However, the trial to trial fluctuations, and more generally
the spontaneous fluctuations of the neural system in question, are not
necessarily   completely   random,  and  quite   often have   explicit
structure,   such  as   oscillations.     Understanding these  "baseline
fluctuations" may be quite important in learning about the dynamics of
neural systems, even in  the context of  response to external stimuli.
To  characterize these spontaneous  fluctuations  is a more  difficult
task  than  learning   input-output relationships.   The    analytical
techniques  developed in this  paper   should be  of  utility  in such
characterization  for multichannel  neural  data, particularly dynamic
brain images.

\section{Different Brain Imaging Techniques}

\subsection{Imaging techniques and their spatiotemporal resolution}

The three main techniques of interest here are optical imaging, 
fMRI and MEG.     Optical imaging     falls   into the    further
sub-categories of  intrinsic   and extrinsic contrast.   In  extrinsic
contrast  optical imaging,  an  optical  contrast agent sensitive   to
neuronal  activity is added  to    the preparation. Examples of   such
contrast agents     include voltage-sensitive  dyes    and  $Ca^{2+}$
concentration  sensitive  contrast    agents. Voltage-sensitive  dye
molecules insert  in the  cell membrane and  the small Stark  shifts
produced in the  molecule by changes  in the transmembrane voltage are
sources  of   the  contrast.  The  signal  to   noise ratio (SNR)   in
these
experiments is typically poor, being of  the order of unity. The 
spatial  
resolution is  set by the optical
resolution and scattering properties of the medium and can be of the  
order of microns. The temporal
resolution  is limited  by  the  digitization  rate of the   recording
apparatus (CCD camera or photodiode array) and  can currently go up to
$\sim 1kHz$, which is the intrinsic timescale of neuronal activity. In
calcium ion sensitive imaging, the   intrinsic timescales are   slower,
so
that the  demands on the digitization rate  are somewhat  less. 
The signal to noise ratio
is  significantly better   compared to  currently available voltage-sensitive 
dyes.  The spatial resolution is greatly enhanced 
in confocal \cite{Pawley} and multi-photon scanning optical 
imaging \cite{Denk90}. The imaging rates in multi-photon 
scanning optical imaging are currently significantly slower than 
the corresponding rates for CCD cameras.

Intrinsic   optical imaging  and  fMRI  rely  on  the same  underlying
mechanism,   namely   hemodynamic   changes  triggered   by   neuronal
activity. Hemodynamic changes include changes in  blood flow and blood
oxygenation level. The  intrinsic timescale for these changes is
slow, ranging from hundreds of  milliseconds to several seconds.   The
intrinsic spatial scale is also  somewhat large, ranging from hundreds
of microns to  millimeters. These scales are well  within the scope of
optical   techniques.  In both  the    extrinsic and intrinsic   cases
discussed  above,   various   noise  sources   including physiological
fluctuations are important indirect determinants of the spatiotemporal
resolution. 

In fMRI,  the instrumental limitations  on  the spatial  and  temporal
resolutions are significant,  and for  fixed  SNR, a tradeoff   exists
between spatial and  temporal resolution, as  well as between temporal
resolution and spatial coverage. The fMRI images are typically gathered
in two  dimensional slices of  finite thickness, and  for a fixed SNR,
the number of  slices  is roughly  linear  in time. For single   slice
experiments, the temporal resolution is $\sim 100 ms$, and the spatial
resolution is $\sim 1 mm$. This spatial resolution can be improved for
a single  slice  by  sacrificing  temporal resolution.    For multiple
slices covering  the whole head,  the temporal resolution is $\sim 2
s$. Note that these numbers  may be expected  to change somewhat based
on future improvements  in instrumentation. The principal advantage of
fMRI is that  it is non-invasive  imaging, making it suitable for  the
study of the human  brain. In addition, optical  imaging is limited to
the  surface  of  the sample,   whereas  fMRI is  a  volumetric imaging
technique. 

In  MEG, weak  magnetic  fields  of the   order  of tens of $fT/\sqrt(Hz)$
generated by  electric currents in the  brain are measured using  
superconducting quantum interfometric detector (SQUID)
arrays positioned   on the  skull.   Like   fMRI, MEG is    a
noninvasive imaging  technique and  therefore applicable to  the human
brain.  The temporal  resolution ($\sim  1ms$) is much  higher than in
fMRI, although  the  spatial resolution  is  in general  significantly
poorer.   The spatial resolution of MEG remains a
debatable issue,  due to the ill-posed  nature of  the inverse problem
that must be solved in order  to obtain an image from  the MEG data. A
promising direction  for future research appears to  be a combined use
of fMRI and MEG, performed separately on the same subject. 

\subsection{Sources of noise}

As mentioned previously, the `noise' present in  the imaging data arise
from two broad  categories of sources, biological  and non-biological.
Biological  noise sources include  cardiac  and respiratory cycles, as
well as  motion  of the  experimental  subject.  In   imaging  studies
involving hemodynamics such as fMRI  and intrinsic optical imaging, an
additional physiological source   of noise are  slow  `vasomotor'  
oscillations \cite{Mitra97}, \cite{Mayhew96}. In
addition in all studies of evoked activity, ongoing brain activity not
locked  to  or   triggered   by the  stimulus   appears   as  `noise'.
Non-biological noise sources  include photon counting noise in optical
imaging experiments, noise in the electronic instrumentation, 60 cycle
noise, building vibrations and the like. 

We first consider optical imaging using  voltage-sensitive dyes in animal
preparations.   These sensitivity of these experiments  is  
currently limited  by  photon
counting noise.  The Stark shifts associated with the available
dyes lead to changes in the optical fluorescence signal on the  order of
$\Delta F/F \sim 10^{-3}/mV$.   The  typical  signal-to-noise ratio is
therefore  of order unity or less.  In  addition, absorption changes
arising from hemodynamic sources,   whether related to the  stimulus or
not, corrupt   the voltage-sensitive dye  images.   Perfusing the brain
with an artificial    oxygen-supplying  fluid   can eliminate    these
artifacts \cite{Prechtl97}.   Motion   artifacts  and  electronic noise  
may  also be
significant.  In contrast, Ca-sensitive dye  images have comparitively 
large signal changes for spike mediated $Ca^{2+}$ fluxes and  are less  
severely  affected by the photon counting
noise.  However motion artifacts can still  be severe, particularly at
higher spatial resolutions. 

In fMRI experiments in   humans,   the instrumental noise   is small,
typically a fraction of a percent.  The dominant noise sources are of
physiological    origin,  mainly cardiac,   respiratory and  vasomotor
sources \cite{Mitra97}.
Depending   on  the time between   successive
images, these oscillations  may  be  well  resolved in the   frequency
spectrum,  or  aliased  and   smeared  over   the sampling   frequency
interval.  Subject motion, due to   respiration or other causes, is  a
major confounding factor in these experiments. 

In MEG experiments the signal-to-noise ratio is usually very low.  
Hundreds of trial averages are sometimes needed to extract evoked 
responses. 
Magnetic fields due to  currents associated with the cardiac cycle
are a  strong noise  source, as  are 60  cycle electrical sources  and
other  non-biological current   sources.    In  studying the    evoked
response, the spontaneous activity not related to the stimulus is a
dominant
source of undesirable fluctuations.

\section{Description of data sets}

In this  section we describe   the data sets  used  to illustrate  the
techniques developed  in the   paper.  We have used  data  from  three
different  imaging   techniques,  corresponding  to  multichannel  MEG
recordings, optical image  time series, and MRI  time series. The data
are grouped into  four  sets, referred to   below using  the  letters \A\ 
through  \D.  Brief  discussions of  the   data collection are provided
below. The data have  either been previously   reported or are gathered
using the same techniques as  in other reports.  In  each case, a more
detailed description may be found in the accompanying reference. 

Data set \A\ consists of multichannel MEG recordings.  Descriptions of
the apparatus and experimental  methods for these data  can be found in
\cite{Joliot94}. For our  purposes it is sufficient  to note  that the
data    were  gathered   simultaneously  from 74     channels  using a
digitization rate of $2.083 kHz$ for a total duration of $5$ minutes, 
and corresponds to magnetic fields due
to spontaneous brain activity recorded from an awake human subject
in a resting state with eyes closed.

Data set \B\ consists of dynamic optical images of the procerebral lobe
of the terrestrial mollusc  Limax  \cite{Kleinfeld94}, gathered  after
staining  the  lobe with a
voltage-sensitive dye.  The digitization rate is $75 Hz$ and the total
duration of the recording is $23 s$. The images are $105 \times  34$ 
pixels in extent and
cover an area of approximately $ 600 \mu m\times 200 \mu m$. 

Data sets \C\ and \D\ contain fMRI data, and consist of time series of
magnetic  resonance images of the human  brain showing a coronal slice
towards  the occipital pole \cite{Mitra97},  \cite{Le96}.   The data
were
gathered in the presence ( \C\ ) or absence ( \D\ ) of a visual
stimulus. 
For data set \C, binocular visual stimulus was provided by a pair
of flickering red LED patterns ($8 Hz$), presented for $30$ seconds 
starting $40$ seconds after the beginning of image acquisition.
The digitization rate for the images
was   $5Hz$ and the total  duration  $110s$. The  images are
$64\times 64$ pixels and cover a field of view of $20 cm\times 20 cm$. 

\section{Analysis Techniques}

\subsection{Time Series Analysis Techniques}

Here  we briefly   review some  methods  used to  analyze time  series
data. The aim here  is not to provide a  complete list of the relevant
techniques, but  discuss those methods  which are directly relevant to
the present  work.  In particular, in the  next  section, we provide a
review of multitaper  spectral  analysis techniques \cite{Thomson82}, 
\cite{Percival93}, since  these are used in the present paper, and are 
not widely known. 

The basic  example to  be considered  is the  power spectral  analysis  
of a
single (scalar) time series, or an output  scalar time series given an
input scalar  time  series. The relevant  analysis  techniques  can be
generally  categorized   under  two different   attributes:  linear or
non-linear, and  parametric   or non-parametric.  We   will mostly  be
concerned   with   multitaper  spectral techniques,  for   which the
attributes are linear and non-parametric.   Although we categorize the
techniques as linear, note that spectra are quadratic functions of the
data.   Also, some of the  spectral  quantities we  consider are other
nonlinear functions of  the  data.  We prefer  non-parametric spectral
techniques (eg.  multitaper spectral estimates) over parametric ones
(eg. autoregressive spectral estimates, also  known as maximum entropy
spectral  estimates, or  linear predictive spectral  estimates).  Some
weaknesses of parametric methods  in the present  context are lack  of
robustness and  lack  of  sufficient flexibility  to   fit data with 
complex spectral content. The reader is referred to  the literature for 
a comparison between  parametric and   non-parametric spectral methods 
\cite{Thomson82}, \cite{Percival93}, and we also discuss this issue 
further below.

We also do not use
methods, based   on time lag  or  delay embeddings,  that characterize
neurobiological  time  series    as outputs   of underlying  nonlinear
dynamical systems. These methods work if the underlying dynamical 
system is low dimensional, and if one can obtain large volumes of data
so as to enable construction of the attractor in phase space. The 
amount of data needed grows exponentially with the dimension of the 
underlying attractor. On the one hand, it is true that neurobiological
time series are outputs of rather nonlinear dynamical systems. However, 
in most cases it is not clear that the constraint of low dimensionality 
is met, except perhaps for very small networks of neurons. In cases 
where the dynamics may appear low dimensional for a short time, 
non-stationarity is a serious issue, and precludes acquisition 
of very long stretches of data. One might think that non-stationarity 
could be accounted for by simply including more dynamical degrees of 
freedom. However, that also would require the acquisition of
exponentially 
larger data sets. We constrain ourselves to spectral analysis techniques
as opposed to the techniques indicated above. The reason  for this is 
twofold:  Firstly, spectral
analysis remains a fundamental  component  of the repertoire of  tools
applied  to  these  problems, and as   far  as  we are  concerned, the
appropriate spectral   techniques  have  not  been sufficiently   well
studied  or utilized  in the   present  context. Secondly, it  remains
debatable whether   much progress has been   made in understanding the
systems involved using the nonlinear techniques \cite{Theiler96},
\cite{Rapp94}. 

\subsubsection{Time domain versus Frequency domain: Resolution
and Non-stationarity}

In the neurobiological context, data are  often characterized in terms
of appropriate correlation  functions. This is equivalent to computing
corresponding spectral quantities.  If  the  underlying processes  are
stationary, then the correlation functions  are diagonal in  frequency
space. For  stationary processes, local  error bars can be  imposed 
for spectra in the frequency domain, whereas the corresponding 
error bars for correlation functions in the time domain are non-local
 \cite{Percival93}.  In addition,  if the  data
contains  oscillatory components, which is  true  for the data treated
here, they are compactly represented in frequency space. These
reasons  form the basis  for using  a  frequency-based representation.
For some other  advantages  of spectra over  autocovariance functions,
see \cite{Percival93} pp.   147-149. The arguments made here are directly 
applicable to the continuous processes that are of interest in the 
current paper. Similar arguments also apply to the computation
of correlation functions for spike trains. An exception should be 
made for those spike train examples where there are sharp features 
in the time domain correlation functions, {\it e.g.} due to 
monosynaptic connections. However, broader 
features in spike train correlation functions would be better 
studied in the frequency domain, a point that is not well 
appreciated.

Despite the advantages of the frequency domain indicated above,
the frequent  presence  of
non-stationarity in the data, makes it necessary  in most cases to use
a time-frequency representation. In  general, the window  for spectral
analysis is  chosen to be  as short as possible  to be consistent with
the spectral structure  of the data,  and this window is translated in
time.     Fundamental  to  time-frequency representations   is  the
uncertainty   principle,  which  sets  the  bounds   for  simultaneous
resolution  in time  and frequency. If   the time-frequency  plane is
`tiled' so as to provide time and frequency  resolutions $\Delta t$ by
$\Delta f$, then $\Delta t \Delta f \ge 1$.  Although there has been a
lot of work   involving tilings of    the time-frequency  plane  using
time-scale   representations (wavelet bases),   we choose to work with
frequency rather than scale   as the basic  quantity, since  the  time
series we are dealing with are better described as having structure in
the frequency domain. In particular, the  spectra typically have large
dynamic   range (which indicates   good  compression  of  data  in the
frequency  space), and also  have   spectral peaks, rather than  being
scale invariant.  The time-frequency analysis is  often crucial to see
this structure, since   a long time average  spectrum  may prove to  be
quite   featureless.   An example of   this  is presented  by MEG data
discussed below.

\subsubsection{Digitization rate, Nyquist Frequency, Fourier Transforms}

Some quantites that are central to the discussion are defined below. 
Consider a time series window of length $T$. The frequency resolution 
is given by the so-called Raleigh frequency, $\delta f = 1/T$. In all 
real examples, the time series is obtained at discrete time locations.
If we assume the discrete time locations are uniformly spaced at 
intervals of $\Delta t$, then the number of time points $N$ in the 
interval $T$ is given by $N=T/(\Delta t)$. The digitization frequency
or digitization rate is by definition $1/(\Delta t)$. An important 
quantity is the Nyquist frequency $f_{Nyq}$, defined as half the 
digitization frequency $f_{Nyq}=1/(2 \Delta t)$. In this context, it 
is important to recall the Nyquist theorem and the related concept
of aliasing. The basic idea of the Nyquist theorem is that the digitized
time series is a faithful representation of the original continuous 
time series as long as the the original time series does not contain any
frequency components above the Nyquist frequency. Put in a different way, a
continuous time series which is band limited to the interval
$[-B,B]$, should be digitized at a rate $2B$ (i.e. $\Delta t=1/(2B)$)
in order to retain all the information present in the continuous 
time series. Aliasing is an undesirable effect that occurs if this 
criterion is violated, namely if a time series with frequency content
outside the frequency interval $[-f_{Nyq},f_{Nyq}]$ 
is digitized at an interval $\Delta t=2/f_{Nyq}$. The spectral power
outside 
the specified interval is then `aliased' back into the interval 
$[-f_{Nyq},f_{Nyq}]$. Consequently the Nyquist theorem tells 
us how frequently a continuous time series should be digitized. 
These concepts are fundamental to the discussion, and the reader 
unfamiliar with them would benifit from consulting an appropriate 
text ({\it e.g.} see \cite{Percival93}).

The Fourier transform $\tilde x(f)$ 
of a discrete time series $\{ x_t | t=n\Delta t, n=1,2,...,N\}$
is defined in this paper as  
 
\begin{equation}
\tilde x(f) = \sum_{n=1}^N x_t e^{-2\pi i f n\Delta t} 
\label{fft_convention}
\end{equation}

\noindent In places where we use the convention $\Delta t =1$
the above equation may be rewritten replacing $n$ by $t$ as 

\begin{equation}
\tilde x(f) = \sum_{t=1}^N x_t e^{-2\pi i f t} 
\end{equation}

The corresponding Nyquist frequency is
dimensionless and numerically equal to $1/2$. In any real application,
of course, both the digitization rate and the Nyquist frequency
have appropriate units.
The total time window length $T$ now becomes interchangeable
with $N$. More generally, $T=N\Delta t$ as above. One frequent 
source of confusion is between the Fourier transform defined 
above and the Fast Fourier Transform (FFT). The FFT is an 
{\it algorithm} to efficiently compute the Fourier transform on
a discrete grid of time points, and should not be confused 
with the Fourier transform which is the underlying continuous
function of frequency defined above.  

\subsubsection{Conventional Spectral Analysis}

In this and subsequent subsections, we set $\Delta t =1$, $T=N$.
Physical units are restored where appropriate.  
We  consider below  a model example  of a  time series constructed  by
adding a  stochastic piece, consisting   of an autoregressive  process
excited by white Gaussian noise, to three sinusoids.  The time 
series is given by $y_t = x_t +
\sum_{k=1}^3 A_k \sin(2\pi   f_k t +\phi_k)$, where  $t=1,2,..,N$ with
N=1024,    and   the   parameters of  the    sinusoids    are  $A_k  =
(0.7,0.7,0.08)$, $f_k = (0.122,0.391,  0.342)$ and $\phi_k = (0,\pi/3,
2\pi/3)$ for $k=(1,2,3)$.  Here  $x_t$ is an autoregressive process of
order 4 given  by $x_t =  \sum_{k=1}^4 a_k x_{t-k}  + \epsilon_t$, 
with $a_k =
(-0.683,1.55,-1.96,1.87)$.  For    successive time      samples   $t$,
$\epsilon_t$ are independently  drawn from a normal distribution  with
unit variance.  In Fig. 2, the first 300 points of the time series
example are plotted. 

\begin{figure}
\centerline{\hbox{\psfig{figure=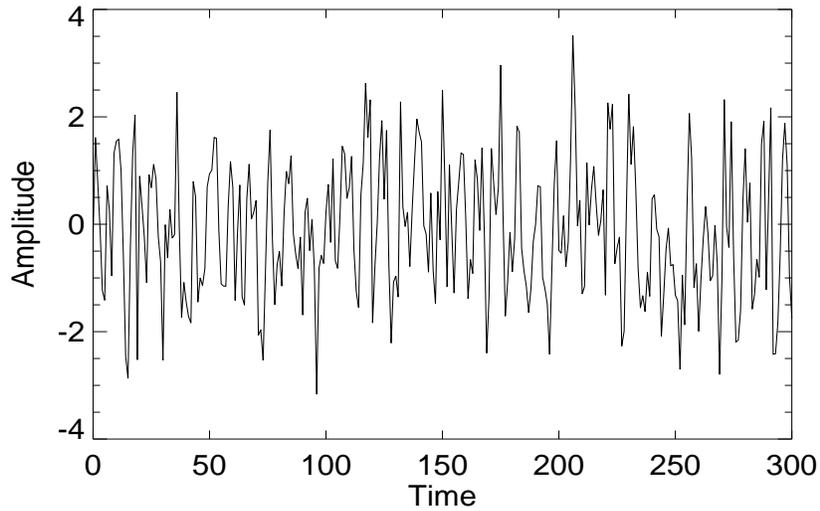,height=3in,width=4.5in}}}
\caption{
Piece of a time series composed of a stochastic piece generated 
by an order 4 autoregressive process added to a sum of three 
sinusoids (see text). 
}
\end{figure}

\begin{figure}
\centerline{\hbox{\psfig{figure=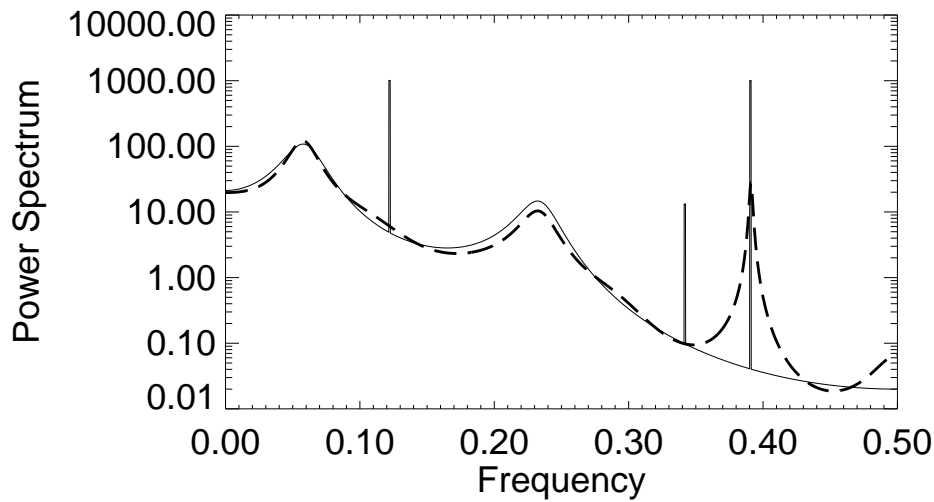,height=3in,width=4.5in}}}
\caption{Autoregressive spectral estimate of time series data shown in 
Figure 2 (dashed line). Also shown is the  theoretical spectrum of the
underlying process (solid line).
Note that the sine waves lead to delta functions, 
whose heights have been determined for the display so that the
integrated intensity over a Raleigh frequency gives the correct strength of 
the theoretical delta function. The autoregressive estimate 
completely misses the peaks at $f=0.122$ and $f=0.391$.
}
\end{figure}

In  conventional  non-parametric spectral analysis  a tapered
\footnote{We use `taper' rather than `window', because `window' is
used below to label a segment of data used in spectral analysis,
particularly in time-frequency analysis.} Fourier
transform  of the data is used  to estimate the  power spectrum. There
are   various choices of  tapers.  A taper with optimal bandlimiting
properties is the zero$^{th}$   discrete prolate spheroidal
sequence which is described in detail later in the text.  Using this
taper, a  single taper spectral estimate is  given in  the top right
corner of Fig. 5.  In this  and the
following section, the Fourier transforms are implemented using 
an FFT after the time series of length $N$ is 
padded out to length $4 N$ or $8 N$. This still gives a discrete 
representation of the corresponding continuous functions of frequency, 
however the grid is sufficiently fine that the resulting function 
appears to be smooth as a function of frequency.

\subsubsection{Autoregressive spectral estimation}
 
To illustrate a parametric spectral estimate, we show in Fig. 3
the results of an autoregressive (AR) modelling  of the data using an
order
19 AR process.  We used the Levinson-Durbin  procedure for purposes of
this illustration \cite{Percival93} p.397. The order of the AR model was
determined using  the criterion   AIC \cite{Akaike74}.  Although   the
parametric estimate  is  smooth, it  fails to accurately  estimate the
underlying theoretical spectrum.  In particular,  it completely misses
the delta function peaks at $f=0.122$ and $f=0.391$. 

Some comments are in order regarding AR spectral estimates. The 
basic weakness of this method is that it starts with the correlation
function of the data in order to compute the AR coefficients. 
This, however, presupposes the answer, since the correlation 
function is nothing but the Fourier transform of the spectrum - 
if the correlation function were actually known, there would be
no need to estimate the spectrum. In practice, an estimate of 
the correlation function is made from the data, which contains
the same bias problems as in estimating spectra. In fact, in 
obtaining the illustrated fit, we computed the correlation function by 
Fourier transforming a direct multitaper spectral estimate 
(to be described below). Attempts to escape from the circularity
pointed out above usually result in strong model assumptions,
which then lead to misfits in the spectra (for further discussions, 
see \cite{Thomson90}, p.614). Despite these problems, AR methods 
do have some use in spectral estimation, namely to obtain
pre-whitening filters that reduce the dynamic range of the process
and thus helps reduce bias in the final spectral estimate. 
Another valid usage of AR methods is to treat sufficiently 
narrowband signals that can be appropriately modeled by low order AR
processes. 

\subsubsection{Multitaper Spectral Analysis}

Here we present a brief review of multitaper 
estimation \cite{Thomson82}.  This method involves the use of multiple
orthogonal data  tapers, in particular  prolate spheroidal functions,
which provide a local eigenbasis in frequency  space for finite length
data sequences.   A summary  of the advantages
of this technique can be found in Percival and Walden \cite{Percival93},
Chapter 7. 

\begin{figure}
\centerline{\hbox{\psfig{figure=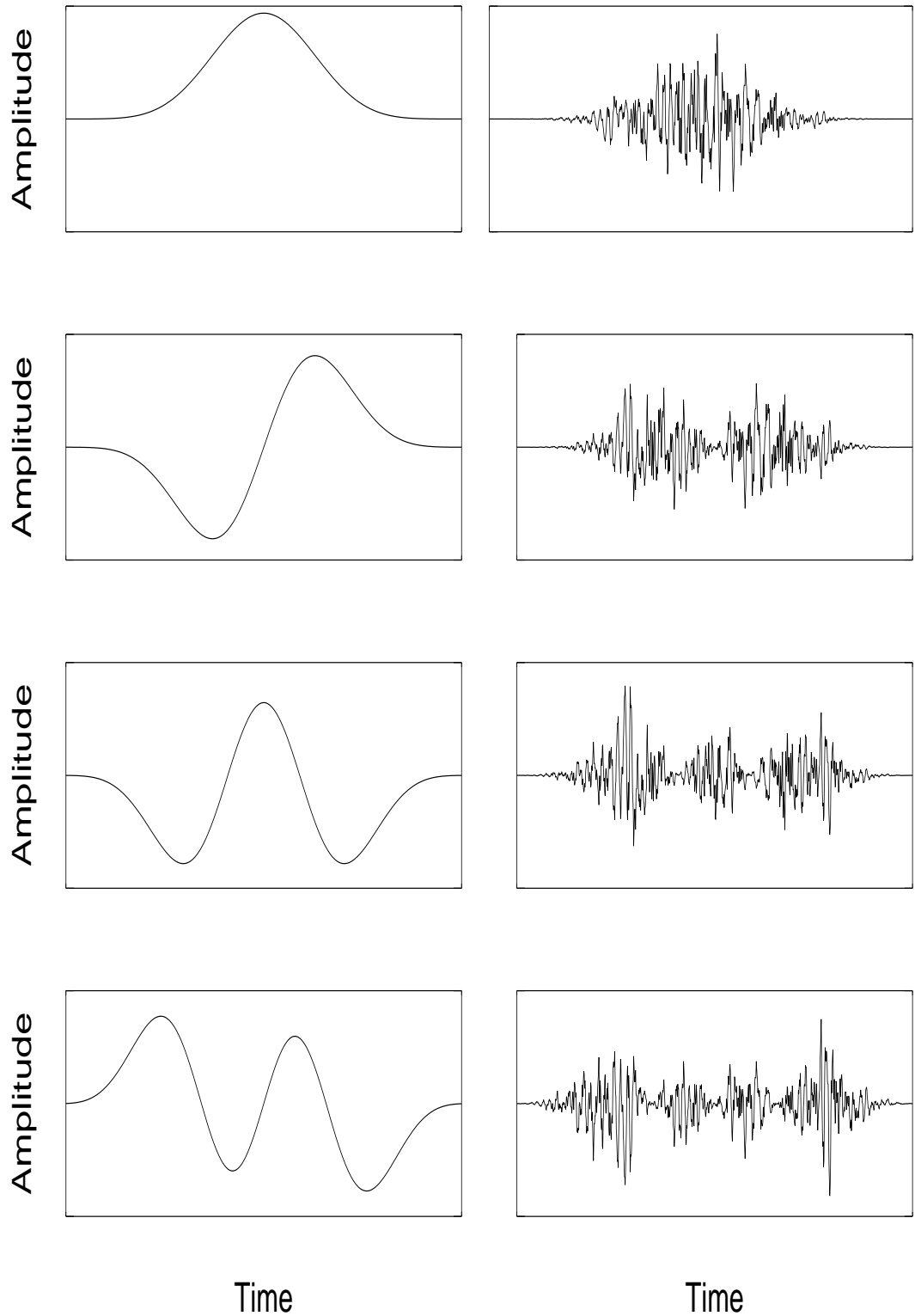,height=8in,width=6in}}}
\caption{
Left column: prolate spheroidal data tapers for $NW=5$
(k=1\ldots4).
Right column: tapered time series corresponding to the time series 
in the example given in the text (part of which is shown in Fig.2),
multiplied by the data tapers in the left column.
}
\end{figure}

\begin{figure}
\centerline{\hbox{\psfig{figure=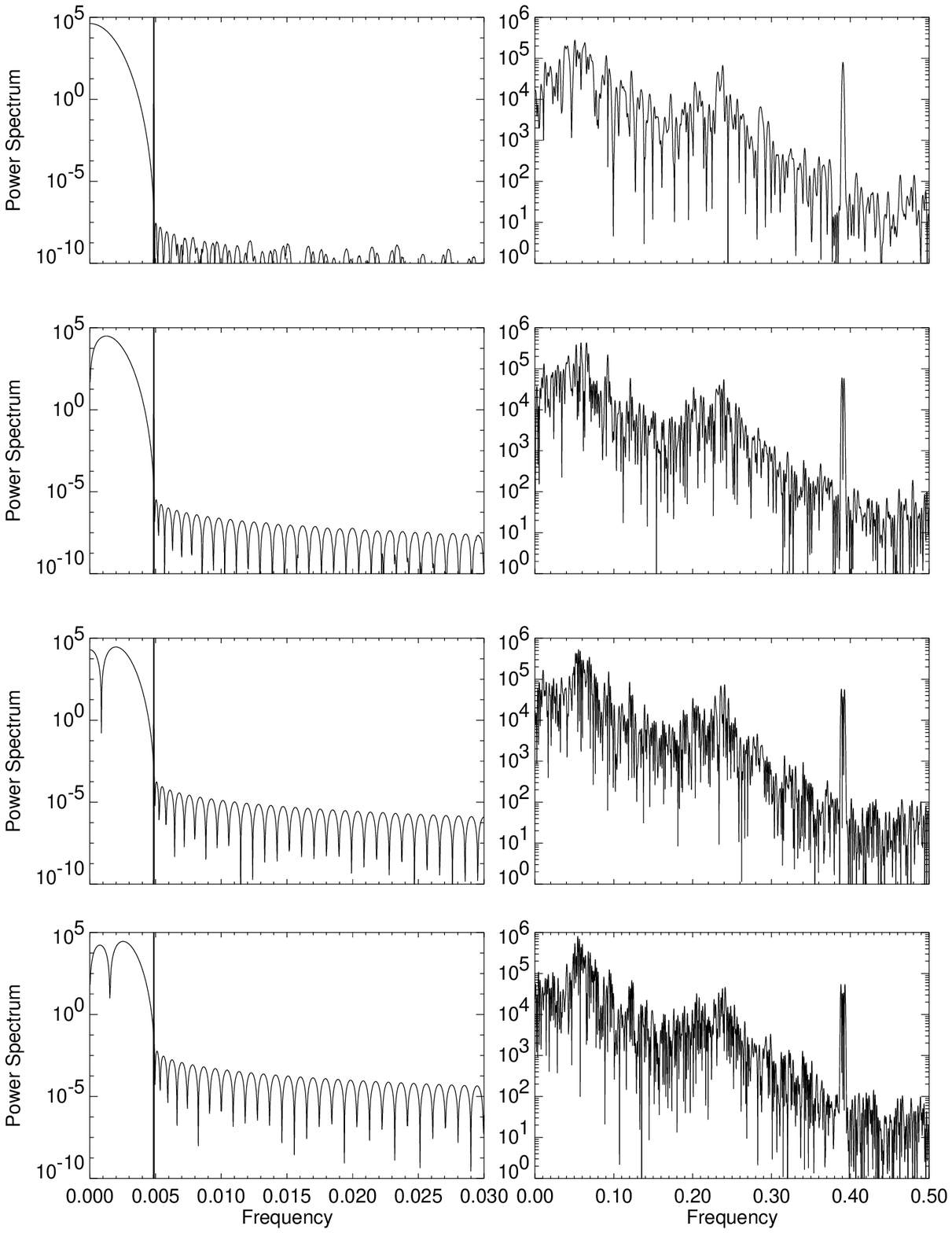,height=8in,width=6in}}}
\caption{
Left column: spectra of the data tapers shown in Fig.4, left column. 
Right column: spectra of the tapered time series shown in Fig.4, right
column, giving single taper estimates of the spectrum. 
}
\end{figure}

Consider  a  finite length sample  of  a discrete  time process $x_t$,
$t=1,2,\ldots,N$.    Let   us assume  a  spectral  representation  for the
process, 
\begin{equation}
x_t = \int_{-\half}^{\half} dX(f) e^{2\pi i f t} 
\end{equation}
The Fourier transform of the  data sequence $\tilde x(f)$ is therefore
given by 
\begin{equation}
\tilde x(f)  = \sum_1^N x_t   e^{-2\pi i f t}  = \int_{-\half}^{\half}
K(f-f^{\prime},N) dX(f^{\prime}) 
\label{spec_convol}
\end{equation}

\noindent where 

\begin{equation}
K(f-f^{\prime},N)   =       e^{-2\pi       i    (f-f^{\prime})(N+1)/2}
{\sin(N\pi(f-f^{\prime})) \over \sin(\pi(f-f^{\prime}))} 
\label{dirichlet_kernel}
\end{equation}

Note that  for   a stationary   process, the   spectrum is given    by
$S(f)df=E[|dX(f)|^2]$.  A simple estimate of  the spectrum (apart from
a normalization   constant)   is obtained   by   squaring the  Fourier
transform of the data  sequence, i.e. $|\tilde x(f)|^2$.  This suffers
from two  difficulties: Firstly, $\tilde x(f)$ is  not equal to $X(f)$
except when the  data length is infinite,  in which case the kernel in
Eq.\ref{dirichlet_kernel} becomes   a delta  function.   Rather, it is
related to $X(f)$ by a convolution,  as given by Eq.\ref{spec_convol}.
This problem is  usually referred to as  `bias', and corresponds  to a
mixing of   information from different  frequencies  of the underlying
process due to a finite data window length. Secondly, even if the data
window length  were  infinite,  calculating $|\tilde  x(f)|^2$ without
using   a tapering function (a  quantity   known as the periodogram)
effectively squares the observations without  averaging - the spectrum
is the {\it expectation} of this squared quantity.  
This issue is referred to as the lack of
consistency  of  the periodogram estimate, namely   the failure of the
periodogram  to converge to the spectrum  for  large data lengths. The
reason for this is  straightforward.  When one takes a fast
Fourier transform of  the data, one is estimating $N$ quantities
from  $N$ data values,  which obviously leads to overfitting  if  the 
data  is stochastic. More precisely, the squared Fourier transform
of the time series is an inconsistent estimator of the spectrum, 
because it does not converge as the data time series tends to infinite
length. 
 
To resolve the first  issue, the data is  usually multiplied by a data
taper, which  leads    to    replacing  the  kernel   in
Eq.\ref{dirichlet_kernel} by a kernel that is more localized in
frequency. However, this leads
to the loss of  the ends of the  data. To surmount the second problem,
the usual approach is to average  together overlapping segments of the
time series \cite{Welch67}.  Repetition  of the experiment also  gives
rise to an  ensemble over which  the expectation can  be taken, but we
are interested  in  single trial   calculations which  involve  only a
single time series.  Evidently, some amount of smoothing is necessary
to reduce the variance of the estimate, the question being what is
an appropriate and systematic way to performing this smoothing.

An elegant approach towards the solution to both of the above problems
is  provided  by    the  multitaper spectral    estimation    method
in which   the data is multiplied  by  not one,  but
several orthogonal  tapers, and Fourier    transformed to obtain  the
basic   quantity for  further spectral   analysis.  The method  can be
motivated by treating Eq.\ref{spec_convol} as  an integral equation to
be solved in a regularized way.  The simplest example of the method is
given by the direct  multitaper spectral estimate $S_{MT}(f)$, defined
as the average over individual tapered spectral estimates, 
 
\begin{equation}
S_{MT}(f) = {1\over K} \sum_{k=1}^K |\tilde x_k(f)|^2 
\end{equation}

where 

\begin{equation}
\tilde x_k(f) = \sum_1^N w_t(k) x_t e^{-2\pi i f t} 
\end{equation}

\begin{figure}
\centerline{\hbox{\psfig{figure=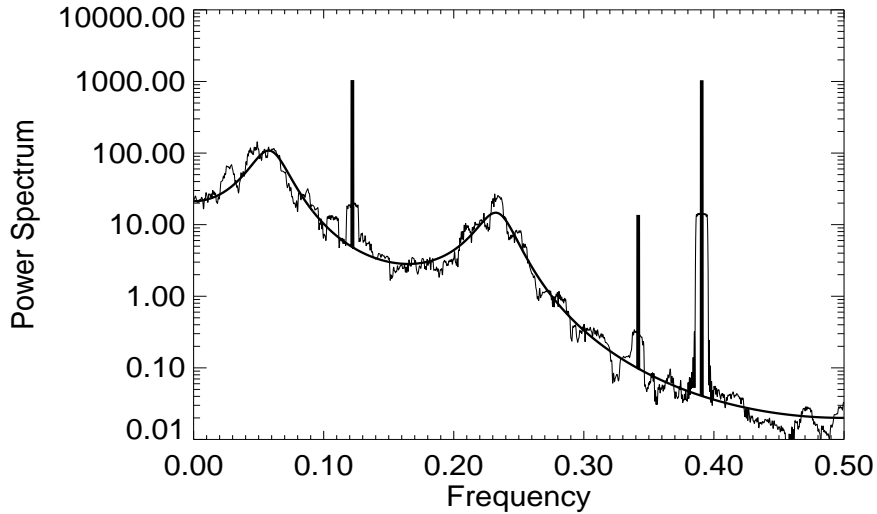,height=3in,width=4.5in}}}
\caption{ 
Direct multitaper spectral estimate of time series example 
using $WT=5$, $K=9$. The estimates for the first four of the nine 
tapers that are averaged to create this estimate is shown in Fig.5.
The theoretical spectrum is also shown.
}
\end{figure}

Here  $w_t(k)~ (k=1,2,\ldots,K)$ are $K$  orthogonal taper functions with
appropriate    properties.   A   particular  choice   of  these taper
functions, with optimal  spectral concentration properties, is  given
by      the    Discrete    Prolate   Spheroidal     Sequences   (DPSS)
\cite{Slepian61}.  Let  $w_t(k,W,N)$ be the $k^{th}$  Discrete Prolate
Spheroidal Sequence (DPSS)    of length $N$   and frequency  bandwidth
parameter    $W$. The  DPSS  form  an  orthogonal  basis  set for
sequences  of  length, $N$,   and are   characterized  by  a bandwidth
parameter $W$.  The important feature of these sequences is that for a
given  bandwidth  parameter $W$   and taper length    $N$, $K=2  NW$
sequences out of a total of $N$ each have  their energy 
effectively concentrated within
a range  $[-W,W]$  of frequency  space. Consider a   sequence $w_t$ of
length $N$ whose Fourier transform  is given by  $U(f) = \sum_1^N  w_t
e^{-2\pi i f t}$. Then  we  can  consider  the  problem  of finding
sequences $w_t$  so that the  spectral  amplitude $U(f)$ is  maximally
concentrated in the interval $[-W,W]$, i.e. 

\begin{equation}
\int_{-W}^W | U(f) |^2 df 
\end{equation}

\noindent is maximized,   subject  to a normalization  constraint   which may be
imposed  using   a Lagrange multiplier.  It  can   be shown  that  the
optimality  condition leads  to   a  matrix  eigenvalue equation   for
$w_t(k,W,N)$ 

\begin{equation}
\sum_{t^{\prime}=1}^N {\sin\bigl[2\pi W (t-t^{\prime})\bigr] \over \pi
 (t-t^{\prime}) } w_{t^{\prime}} = \lambda w_t 
\label{prolate_eqn}
\end{equation}

The eigenvectors of this equation are the DPSS. The remarkable fact is
that the first    $2 NW$   eigenvalues    $\lambda_k(N,W)$ 
(sorted in descending order)  are  each approximately  equal to one, 
while the remainder are approximately  zero.  Since it follows from 
the above definitions that

\begin{equation}
\lambda_k(N,W) = {\int_{-W}^W |U(f)|^2 df \over
\int_{-{1\over 2}}^{1\over 2} |U(f)|^2 df}
\end{equation} 

\noindent this is a precise statement of the spectral concentration mentioned 
above. 
The fact that many of the eigenvalues are close to one
  makes  the eigenvalue problem Eq.\ref{prolate_eqn}  ill-conditioned 
and unsuitable for the actual  computation of the prolates
(this   can  be achieved   by a  better  conditioned  tridiagonal form
\cite{Percival93}).  The DPSS can be shifted in concentration
from $[-W,W]$ centered around  zero  frequency to any non-zero  center
frequency  interval  $[f_0-W,f_0+W]$  by    simply  multiplying   by  
the appropriate phase  factor  $e^{2\pi i  f_0 t}$,   an operation  known as
demodulation. The usual strategy is to
select the desired analysis half-bandwidth  $W$ to be a small multiple
of the Raleigh frequency $1/N$, and then take the  leading $2 NW - 1$
DPSS as data tapers in the multitaper  analysis.  Note that less than
$2  NW$ of the sequences  are typically taken, since  the last few of
these have progressively worsening spectral concentration properties. 

For illustration, in the left column of Fig.  4, we show the first $4$
DPSS  for $W=5/N$. Note  that the orthogonality condition ensures that
successive DPSS each have one   more zero crossing than the   previous
one. In the right column of Fig. 4, we show the time series example
from the earlier subsection multiplied  by each of the successive data
tapers. In the left column of Fig. 5, we show the spectra of the data
tapers themselves, showing   the spectral concentration property.  The
vertical marker denotes the bandwidth parameter $W$ 

In Fig. 5, we show the magnitude squared Fourier transforms of the
tapered time series shown in Fig. 4.  The arithmetic average of these
spectra  for $k=1,2,\ldots,9$ (note that  only  4 out of  9  are shown  in
Figs. 4 and 5) give a direct multitaper  estimate of the underlying process,
shown in Fig.6.  Also shown in that figure is the theoretical spectrum
of the underlying model. 

In the direct  estimate, an arithmetic average of
the different spectra is   taken. However, the different data   tapers
differ in their spectral sidelobes, so that a weighted average is more
appropriate.  In  addition, the   weighting factors should   be chosen
adaptively depending on the local variations in the spectrum. For more
detailed considerations along these lines,  the reader is referred  to
\cite{Thomson82}. 

\begin{figure}[t]
\begin{center}
\mbox{
\makebox[0 in][l]{$(a)$}\psfig{figure=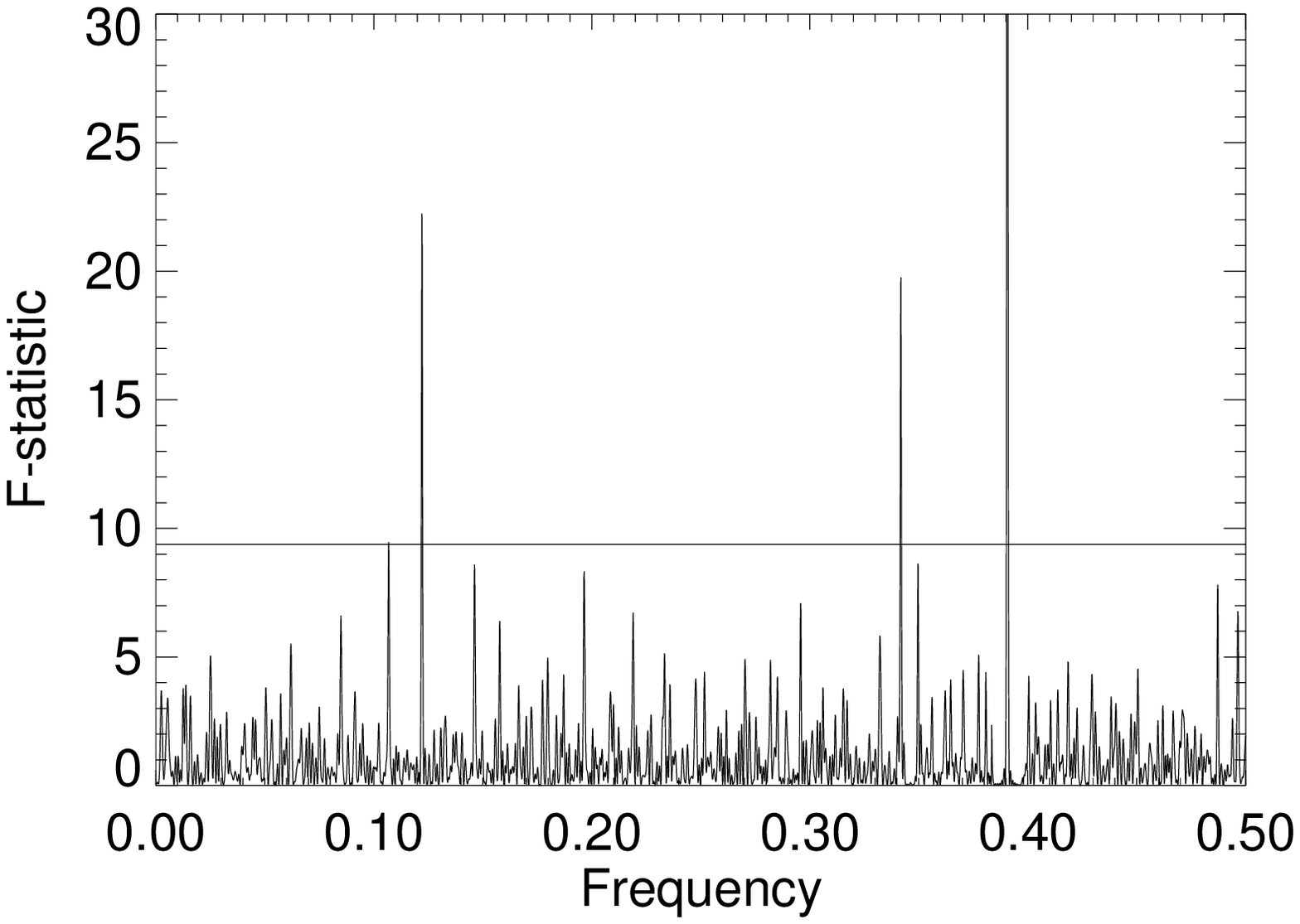,width=3.5 true
in,height=2.5 true in}
\hspace{0.07in} 
\makebox[0 in][l]{$(b)$}\psfig{figure=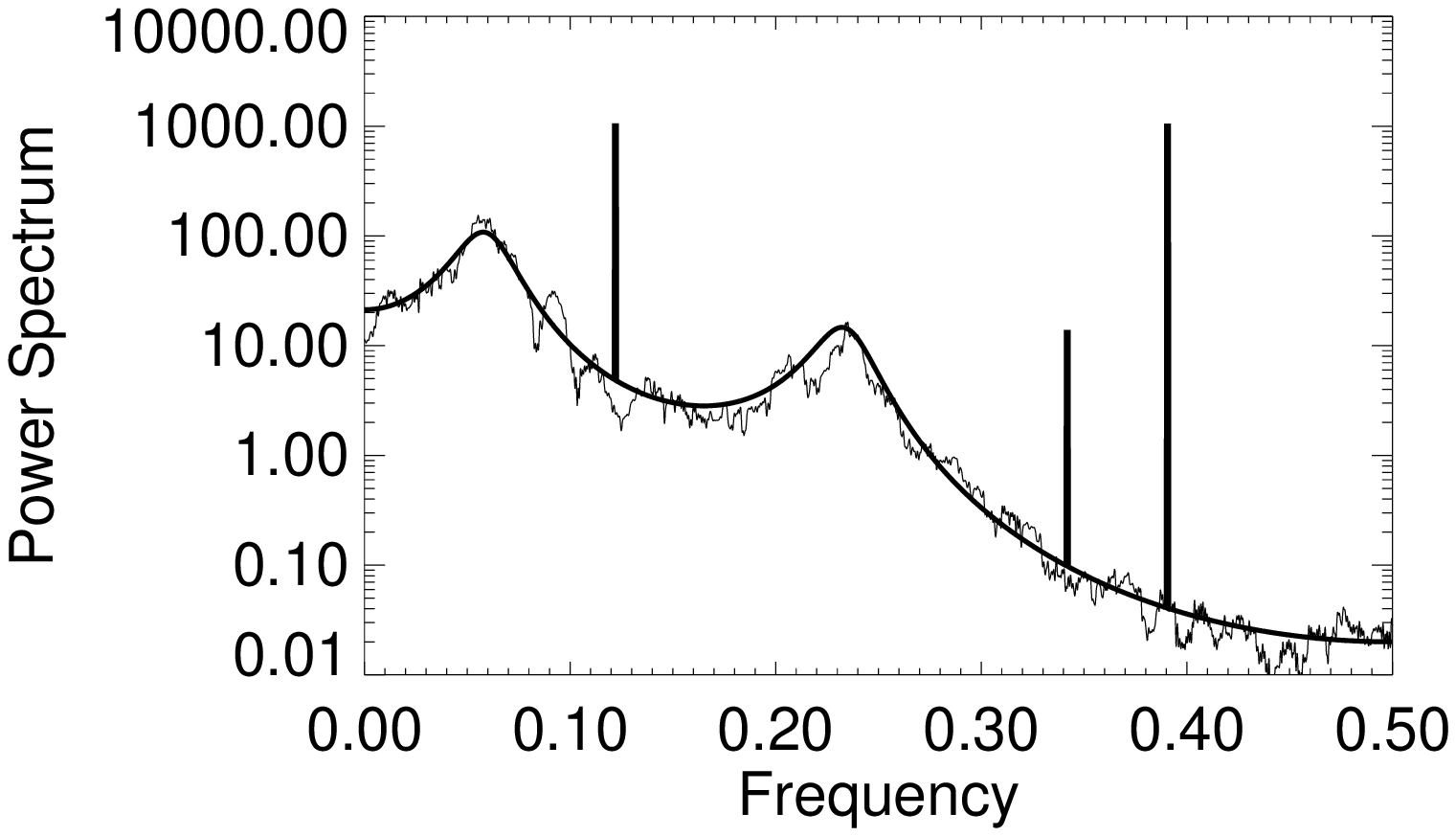,width=3.5 true
in,height=2.5 true in}   
}\\
\end{center}
\caption[abbrev]{ 
{\em a)} F-statistic values for testing for the
presence of a significant sinusoidal component in the time series 
example along with significance levels.  
{\em b)} Direct multitaper spectral estimate of the time series example
using $WT=5$, $K=9$
reshaped to remove the sinusoidal lines.  The theoretical spectrum is
also shown in bold. 
}
\end{figure}

Sine waves in the  original  time series correspond to
square peaks in  the multitaper spectral  estimate. This is usually an
indication that the time series contains  a sinusoidal component along
with a  broad background. The presence of  such a sinusoidal component
may   be detected by  a  test  based on  a goodness-of-fit F-statistic
\cite{Thomson82}.  To proceed, let us  assume that the data contains a
sinusoid of  complex    amplitude $\mu$   at    frequency $f_0$   (The
corresponding real  series being $Re(\mu e^{2\pi i  f_0 t})$).  Let us
also assume that in a frequency interval $[f_0-W,f_0+W]$ around $f_0$,
the process, to which the sinusoid is added, is white.  Note that this
is a much weaker assumption  than demanding that  the process is white
over  the  entire   frequency  range.  Under   these assumptions,  for
$f\epsilon [f_0-W,f_0+W]$ the  tapered Fourier transforms of the data
are given by 

\begin{equation}
\tilde x_k(f) = \mu U_k(f-f_0) + n_k(f), ~~~ k=1,2,\ldots,K 
\label{line_regression}
\end{equation}

Here $U_k(f)$ is the Fourier transform of the $k^{th}$  DPSS, and f is
 in  the  range $[f_0-W,f_0+W]$.  The  assumption of  a  locally white
 background implies that $n_k(f)$  are independently and indentically 
 distributed complex Gaussian variables.
 Treating Eq.\ref{line_regression} as  a linear regression equation at
 $f=f_0$   leads to  an  estimate of   the sine  wave amplitude (which
 corresponds to a particular tapered Fourier transform of the data) 

\begin{equation}
\hat{\mu}(f_0) = {\sum_k x_k(f_0) U_k^*(0) \over \sum_k | U_k(0) |^2} 
\label{line_estimate}
\end{equation}

\noindent and to an $F$ statistic for the significance of a non-zero $\mu$ 

\begin{equation}
F(f_0) =   { |\hat{\mu}(f_0)|^2  \over  \sum_k   |  \tilde  x_k(f_0) -
\hat{\mu}(f_0) U_k(0) | } 
\end{equation}

Under the null hypothesis that there is  no line present, $F(f_0)$ has
an $F$ distribution with  $(2,2K-2)$ degrees  of  freedom. We  plot this
function   $F(f)$  for  the   time series in    the above   example in
Fig. 7(a). For this example, we have chosen $W = 7/N$, $K=13$.  One
obtains an  independent  $F$  statistic every Raleigh  frequency,  and
since   there  are  $N$ Raleigh  frequencies    in  the spectrum,  the
statistical  significance level is chosen   to be $1-1/N$. This  means
that on  an average, there  will be at most one  false detection  of a
sinusoid across all frequencies.  A horizontal line in Fig. 7(a)
indicates this significance level. Thus, a line crossing this level is
found to be a significant sinusoid present in the data.  The sinusoids
known to be  present in  the  data are  shown to   give  rise to  very
significant  F-statistics at this  level of significance, and there is
one spurious crossing of the threshhold by a small amount.  The linear
regression leads to an  estimate for the  sinusoid amplitudes. In  the
present example, the percentage   difference between the  original and
estimated amplitudes were  found to  be 6\%,  4\%, 2\% for  increasing
frequencies. The  corresponding errors for  the phase were 0.2\%, 2\%,
2\%.   Note that the errors in  phase estimation  are smaller than the
errors in the amplitude estimates.   From the estimated amplitudes and
phases, the sinusoidal  components can be reconstructed and subtracted
from the  data. The spectrum  of the residual  time series can then be
estimated by the  same techniques. This residual  spectrum is shown in
Fig. 7(b), along with the theoretical spectrum of the underlying
autoregressive process. 

\subsubsection{Choice of the bandwidth parameter}

The choice of the time window length $T=N\Delta t$ and the bandwidth 
parameter $W$ is critical for applications. No simple procedure can be
given 
for these choices, because the choice really depends on the data set
at hand, and is best made iteratively by visual inspection and some 
degree of trial and error. $2TW = 2W/(1/T)$ gives the number
of Raleigh frequencies over which the spectral estimate is effectively 
smoothed, so that the variance in the estimate is typically reduced 
by $2TW$. Thus, the choice of $W$ is a choice of how much to smooth. 
  In  qualitative terms,
the bandwidth parameter should be chosen to  reduce variance while not
overly distorting  the spectrum. This can be  done formally by trading
off an appropriate weighted sum of the estimated variance and bias.
However, as a rule of the thumb we find 
fixing the time bandwidth product $TW$ at a small number 
(typically 3 or 4) and then varying the window length in time until 
sufficient spectral resolution is obtained is a reasonable strategy.
This presupposes that the data is examined 
in the time frequency plane, so that $T$ may be significantly smaller 
than the total data length.  

\subsection{Analysis of Multivariate Data}

So far   we have  concentrated on   the analysis  of   univariate time
series.  However, the principal subject of  this paper is multichannel
data, so we now consider the analysis of multivariate time series. The
basic methods for  dealing with such  data are  similar to those 
used for other
multivariate data, with modifications  to  take into account the  fact
that we are dealing with  time series. In  fact, a scalar time  series
can itself  be usefully represented  as a multivariate time  series by
going     to a   lag-vector    representation,  or  a   time-frequency
representation. Such a  representation may be  desirable to understand
the structure  underlying the  scalar  time series.  Other examples of
multichannel data include multiple spike trains,  and various forms of
brain imaging, including optical imaging using intrinsic and extrinsic
contrast     agents,        magnetic      resonance     imaging    and
magnetoencephalography. In  general one can  think of one or two space
dimensions added to  the time dimension in  the data. In  the sections
below, we review briefly  some of the concepts  useful to our analysis
later in the  paper. The techniques  can  be grouped into  two general
classes:  choosing   an  appropriate  low   dimensional representation
(e.g. in Principal Components Analysis) or  choosing a partitioning of
the multidimensional space (e.g. in  various forms of clustering). We
concentrate in this   section on mode  decomposition.  A discussion of
clustering methods may be found in \cite{Duda}. 

\subsubsection{Eigenmode Analysis: SVD}

The   Singular Value Decomposition   (SVD)  is a  representation of  a
general  matrix of fundamental importance   in linear algebra that  is
widely   used to  generate  canonical  representations of multivariate
data. It is equivalent to Principal Component Analysis in multivariate
statistics, but  in addition  is  used  to generate  low   dimensional
representations for complex multidimensional time  series. The SVD of
an  arbitrary (in general complex) $p\times  q$ matrix ($p > q$) 
${\cal M}$ is given by ${\cal M} = U \Lambda  V^{\dagger}$, where the
$p\times
q$ matrix  $U$ has orthonormal columns, the  $q\times q$ matrix
$\Lambda$
is  diagonal with real, non-negative  entries  and  the  $q \times  q$
matrix $V$ is    unitary.  Note  that  the matrices   ${\cal M}  {\cal
M}^{\dagger} = U  \Lambda^2 U^{\dagger}$ and ${\cal M}^{\dagger} {\cal
M}  = V   \Lambda^2   V^{\dagger}$  are hermitian,  with   eigenvalues
corresponding to the diagonal  entries of $\Lambda^2$  and U and V the
corresponding matrices of  eigenvectors.  Consider the special case of
space-time data $I({\bf x}, t)$.  The SVD of such data is given by 

\begin{equation}
I({\bf x} , t) = \sum_n \lambda_n I_n({\bf x}) a_n(t) 
\end{equation} 

\noindent where $I_n({\bf  x})$ are the  eigenmodes of the "spatial correlation
matrix" 

\begin{equation}
C({\bf   x},   {\bf   x}^{\prime})  =   \sum_t  I({\bf    x},t) I({\bf
x}^{\prime},t) 
\end{equation}

Similarly $a_n(t)$  are the eigenmodes   of the "temporal  correlation
function" 

\begin{equation}
C(t, t^{\prime}) = \sum_{\bf x} I({\bf x},t) I({\bf x},t^{\prime}) 
\end{equation}

If the sequence of  images were randomly   chosen from an ensemble  of
spatial images, then $C({\bf x},  {\bf x}^{\prime})$ would converge to
the ensemble  spatial correlation function  in the limit  of many time
samples.   If  in  addition  the  ensemble  had   space  translational
invariance  then  the eigenmodes  $I_n({\bf x})$ would  be plane waves
$e^{i{\bf k}\cdot  {\bf x}}$, the mode number  "n" would correspond to
wavevectors  and the singular values   would correspond to the spatial
structure  factor $S({\bf k})$.    In  general, the image ensemble  in
question will not have translational invariance;  however the SVD will
then  provide a basis set  analogous to wave  vectors.  In physics one
normally  encounters structure  factors  $S({\bf k})$ that decay  with
wave vector. In  the more general  case, the singular  value spectrum,
organized  in descending order, will   show a decay  indicative of the
structure in the data. 

\begin{figure}
\centerline{\hbox{\psfig{figure=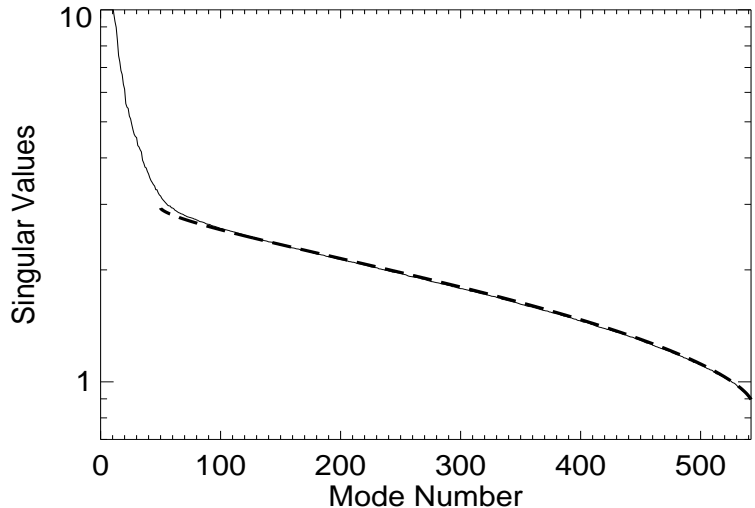,height=3in,width=4.5in}}}
\caption{
Sorted singular values corresponding to a space-time SVD of 
functional MRI image time series from data set \D . 
The tail in the spectrum is fit 
with the theoretical formula given in the text.
}
\end{figure}

To make sense of an SVD performed on a data matrix, it is important to
know the expected  distribution of singular values  if  the entries of
the  matrix were  random. This  problem lies in  the  domain of random
matrix  theory, and can be solved  in special cases (Sengupta and
Mitra, 1996, Unpublished).
As an example, consider  the case of a $p\times  q$ matrix ${\cal M} =
{\cal M}_0 + N$ where ${\cal M}_0$ is fixed and the entries of $N$ are
independently  normally  distributed  with  zero   mean and  identical
variance $\sigma^2$.  ${\cal M}_0$ may be thought of as the desired or
underlying signal;  for  an  SVD  to be  useful,  ${\cal M}_0$  should
effectively have a low rank structure. A typical  procedure is to take
the SVD of ${\cal M} $ and to  truncate the singular value spectrum to
keep only values that cross a threshhold. Consider the special case in
which ${\cal  M}_0= 0$, so  that  we are dealing  with a  purely noise
matrix. In   this    case, it can   be   shown  that \cite{Denby91},
(Sengupta and Mitra, 1996, Unpublished)  the density of singular values,
defined as 

\begin{equation}
\rho(\lambda) =\langle \sum_n \delta(\lambda-\lambda_n) \rangle 
\end{equation}

\noindent is given in the limit of large matrix sizes by 

\begin{equation}
\rho(\lambda)          =          {1\over            \pi\sigma\lambda}
\sqrt{(\lambda_+^2-\lambda^2) (\lambda^2-\lambda_-^2)} 
\label{svdist}
\end{equation}

\noindent where  (Recall that $\sigma^2$ is the  variance of the matrix entries,
and $p,q$ are the dimensions of the matrix) 

\begin{equation}
\lambda_{\pm}^2 = 2\sigma^2 \Bigl[{p+q\over 2} \pm \sqrt{p q}\Bigr] 
\end{equation}

It is somewhat easier  to work with  the integrated density  of states
$P(\lambda)  = \int_0^\lambda \rho(x)  dx  $, since $\lambda $ plotted
against $1-P(\lambda)$ gives the sorted singular values (in decreasing
order).  More generally, ${\cal M}_0  $ is not zero but  is given by a
low  rank matrix.   The distribution  of  the singular   values can be
worked out in this case, but  if the original matrix  has low rank and
if the `signal'  singular values  are large  compared to  the  `noise'
singular  values, then the   singular value distribution  shows a tail
which  can be fit by  the  above formula,   the only quantity  needing
adjustment     being the total    weight  under  the density  function
$\rho(\lambda)$ (i.e. an overall normalization factor). 

To illustrate  the above, we show  in Fig.8 the sorted singular values
from the SVD of data set \C\ consisting of 550 frames of fMRI data.  The
original image data consisted of 550 images each $64 \times 64$ pixels
big, but
the space data was  first masked to  select out 1877 pixels. Thus, the
SVD is   performed on  a $1877  \times   550 $ matrix.   The resulting
singular values   are shown with  the range  truncated  to
magnify the `noise' tail. Also shown is the theoretical singular value
spectrum expected for  the  noise  tail (dashed  line)  based  on  the
formula  Eq.\ref{svdist} for a    pure noise  matrix, with  a   single
adjustable parameter $\sigma^2$ which has  been selected to match  the
middle portion of  the the tail. The total  weight of the density  has
been adjusted to account for the last 500 singular values. 

\setcounter{figure}{8}
\begin{figure}[t]
\makebox[0 in][l]{$(a)$}\psfig{figure=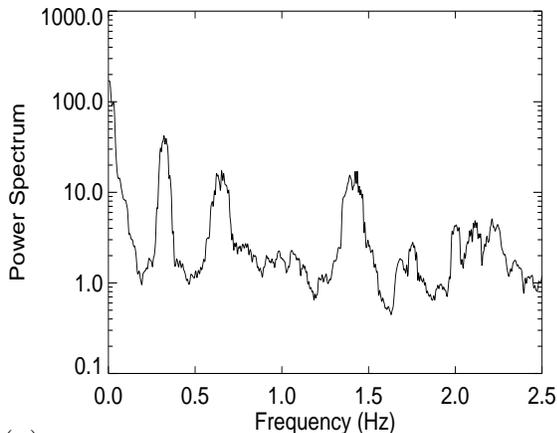,width=3.0 true
in,height=2.53 true in}
\hspace{0.1in} 
\makebox[0 in][l]{$(b)$}\psfig{figure=figure9b.ps,width=3.0 true
in,height=3.1 true in}   
\caption{
Spectra of first 20 principle component time series from data set
\C . 
{\em a)} Weighted Average Spectrum. {\em b)} Image showing the spectra
versus
mode number. The log amplitudes of the spectra are  color coded.
}
\end{figure}

Unlike  in  the  temporal domain,  where  going  to a  frequency-based
representation  does make sense for  neurobiological data, the spatial
wave-vector representation  is   not of  general  use because   of the
generic lack   of  translational  invariance in   space.  However, the
spatial basis generated by an SVD is somewhat more meaningful. It may,
for example, reflect underlying  anatomical structure.  Application of
the SVD on space-time  imaging data  may  be found in  the literature,
sometimes with modifications. However, the space-time SVD suffers from
a severe drawback in the present context. The difficulty is that there
is   no reason why  the  neurobiologically distinct  modes in the data
should be  orthogonal to  each   other, a constraint imposed   by  the
SVD. In   practice, it is  observed that  an  SVD  on space-time data,
different  sources of fluctuations,  such  as cardiac and  respiratory
oscillations and results of  neuronal activity may  appear in the same
mode of the decomposition, thus preventing an effective segregation of
the different degrees of freedom. 

As  an  example, consider  the  SVD   of the  fMRI  data   set \C\  the
corresponding  singular value distribution of which  has been shown in
Fig.8. Note that  in this data, the digitization  rate was $5 Hz$, and
the length of  the time series $110 s$.  The mixing of physiologically
distinct processes in  the individual principal component time  series
thus obtained can  be  seen by studying the  spectra  of the principal
components. 
Figure 9(a) shows the average spectrum across principal components of
data set \C .  
The peak near
zero  frequency  corresponds to the stimulus response as well as possible vasomotor oscillation or other slow flucutations.  The peaks at
$0.3Hz,  0.6Hz$   correspond to  breathing,  and  the  peak at $1.0Hz$
corresponds to  the  cardiac  cycle. 
In Fig.9(b), the spectra are  shown for the first $20$ modes
(with the largest   singular values). The   spectra are coded  as grey
scale intensities,   and are   shown against  the  corresponding  mode
numbers.  
As is   clear  from  studying the
spectra as  a function  of  mode number Fig.9(b),  the decomposition
mixes the various effects. 

We  describe below  a   more   effective way of   separating  distinct
components in the  image    time series  dealt   with  here  using   a
decomposition   analogous  to    the  space-time   SVD,   but in   the
space-frequency domain. The success of  the method stems from the fact
that the data in question is better characterized by a frequency based
representation. 

\subsubsection{Space-Frequency SVD}

From  the  presence of   clear spectral peaks  in  Fig.9(a)  it can be
inferred that different  components in the  dynamic image  data may be
separated if the SVD were  performed after localizing  the data in the
frequency domain.   This can be achieved  by projecting the space-time
data  to  a  frequency   interval, and  then   performing SVD  on this
space-frequency data \cite{Thomson91}, \cite{Mann94}, \cite{Mitra97}. 
Projecting the data on a frequency
interval  can  be   performed   effectively by using   DPSS   with the
appropriate bandwidth parameter. For a  fixed center frequency $f$ and
a half bandwidth $W$, consider the projection matrix 

\begin{equation}
P_{k,t}(f;W)=w_k(t,W) e^{2\pi i t f} ~~~ k=1,2,..,K 
\label{proj_op}
\end{equation}

In  the above, we assume  that  $K =  [2NW]$, where $[X]$ is  the
integer closest to  X but  less  than X. Consider the space of
sequences of length $N$, this  matrix  projects out  a
subspace  with frequencies
concentrated  in $[f-W,f+W]$. Note that  $P^{\dagger}  P$ serves as an
optimal bandlimiting filter on the time series.  Given the $N_x \times
N$ space-time data matrix  $I  = I(x,t)$, the space-frequency   data
corresponding  to the frequency band $[f-W,f+W]$  is given by the $N_x
\times K$ complex matrix $\tilde I = I P^T$. In expanded form, 

\begin{equation}
\tilde I(x,k; f) = \sum_{t=1}^N I(x,t) w_k(t,W) e^{2\pi i t f} 
\end{equation}

We are  considering here  the  SVD of the  the  $N_x \times K$ complex
matrix with entries $\tilde I(x,k; f)$ for fixed $f$. 

\begin{equation}
\tilde I(x,k; f) = \sum_n \lambda_n(f) \tilde I_n(x; f) a_n(k;f) 
\end{equation}

This SVD can be carried out as a function of the center frequency $f$,
using an   appropriate choice of $W$.  In   the best case  most of the
coherent structure is captured by the dominant singular vector at each
frequency.  At each frequency f, one obtains a singular value spectrum
$\lambda_n(f)$  (n=1,2,..,K), the  corresponding (in general  complex)
spatial mode $\tilde I_n(x;f)$, and  the corresponding local frequency
modes $\tilde a_n(k;f)$.   The frequency modes  can  be projected back
into  the   time domain using $P_{k,t}(f;W)$  to   give (narrowband)
time varying
amplitudes of the complex eigenimage. 

In the space-frequency SVD  computation,  an overall coherence  
$\bar{C}(f)$ may be
defined as (it is assumed that $K \leq N_x$)

\begin{equation}
\bar{C}(f) = {\lambda^2_1(f) \over \sum_{n=1}^K \lambda^2_n(f) } 
\end{equation}

The overall coherence spectrum then reflects  how much of the 
fluctuations  in
the  frequency band  $[f-W,f+W]$  is captured  by the dominant spatial
mode. If the image data is completely coherent in that frequency band,
then $\bar{C}(f)=1$. More generally, $1\ge \bar{C}(f) \ge 0$,
 and for random data,
assuming $N_x \gg  K$, $\bar{C}(f) \sim {1\over K}$.
If $N_x$ and $K$  are
comparable, then results such as those  in the previous section may be
used to determine the distribution of $\bar{C}(f)$. 

\setcounter{figure}{9} 
\begin{figure}
\centerline{\hbox{\psfig{figure=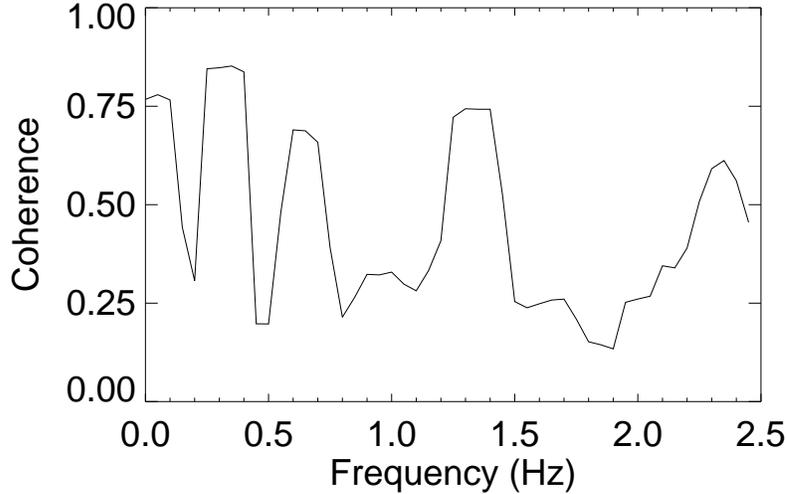,height=3in,width=4.5in}}}
\caption{
Overall coherence spectrum corresponding to the functional 
MRI time series examined in Figs.8 and 9. 
}
\end{figure}

To illustrate  this technique, we show results of its
application to data set \C\ of fMRI  data. The calculation used 19 DPSS,
corresponding to a full bandwidth  of $0.2Hz$.  The overall coherence 
spectrum
resulting from a space-frequency SVD analysis of this data is shown in
Fig.10.   Note  the  correspondence  of     this  spectrum,  which  is
dimensionless,  to the  power  spectrum presented   in  Fig.9(a).  The
magnitudes $|\tilde I_1(x;f)|$ of the dominant spatial eigenmodes as a
function of frequency  are  shown in Fig.11.   The  leading eigenmodes
separate  the   distinct sources of   fluctuations as   a  function of
frequency. 

\setcounter{figure}{10}
\begin{figure}
\centerline{\hbox{\psfig{figure=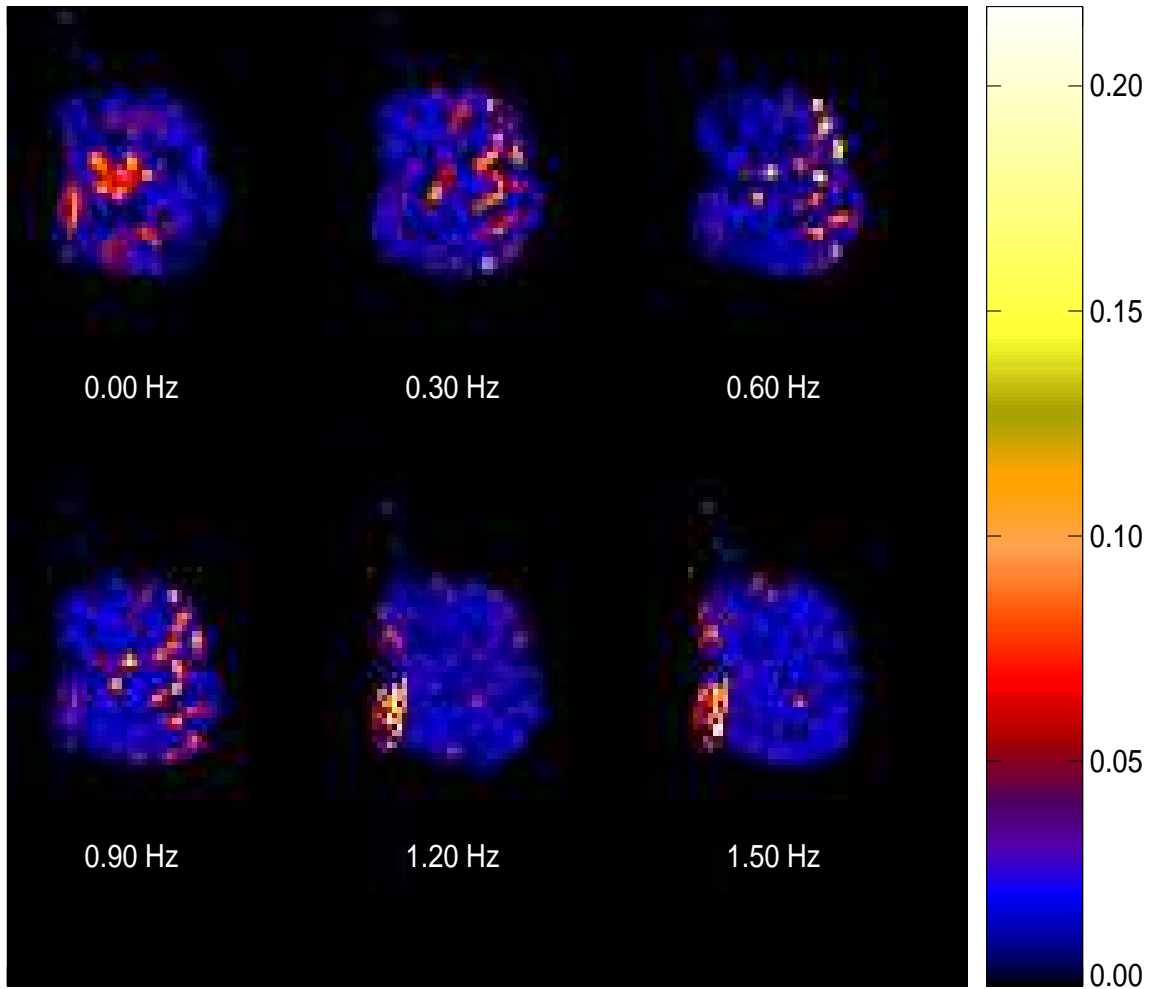,height=6in,width=6in}}}
\caption{
Amplitudes of leading spatial eigenmodes for the space-frequency SVD
of fMRI datai from data set \C .
Note that the spatial structure varies as a function
of center frequency with the physiological oscillations segregating into
distinct frequency bands.
}
\end{figure}

\subsection{Local Frequency Ensemble and Jackknife Error Bars}

One important advantage  of the multitaper method  is that it offers a
natural way of estimating error  bars corresponding to most quantities
obtained  in time  series  analysis, even if  one  is  dealing with an
individual instance of a time series.   The fundamental notion here is
that  of a {\it   local frequency ensemble}. The  idea  is that if the
spectrum of  the process is locally flat  over a bandwidth  $2W$, then
the tapered Fourier transforms $\tilde x_k(f) = \sum_{t=1}^N e^{-2\pi
i f t} x(t) w_k(t)$ constitute a  statistical ensemble for the Fourier
transform of  the process  at   the frequency $f_0$. Assuming  that  the
underlying     process is  locally    white   in the  frequency  range
$[f_0-W,f_0+W]$, then it follows from the orthogonality of the data
tapers
that $\tilde x_k(f)$  are uncorrelated random  variables with the same
variance.   For large N,    $\tilde  x_k(f)$ may  be   assumed to   be
asymptotically  normally distributed  under some general circumstances
(For related results see  \cite{Mallows67}). This provides one way  of
thinking  about  the  direct multitaper   estimate  presented in   the
previous sections: the estimate consists of  an average over the local
frequency ensemble. 

The  above discussion serves  as a motivation for multitaper estimates
of the correlation function,  the transfer function and  the coherence
between two time  series.  Given two time series  $x_t,  y_t$, and the
corresponding   multiple tapered  Fourier  transforms $\tilde x_k(f),
\tilde y_k(f)$, the    following   direct estimates  can   be  defined
\cite{Thomson82}  for the correlation  function  $C_{yx}(f) = E[\tilde
y(f) \tilde x^*(f) ]$,  the  transfer function $T_{yx}(f) =   E[\tilde
y(f) \tilde  x^*(f) ]/E[|\tilde x(f)|^2 ]$  and the coherence function
$\rho_{yx}(f)  =  E[\tilde y(f)  \tilde   x^*(f)  ] /  \sqrt{E[|\tilde
x(f)|^2 ]E[|\tilde y(f)|^2 ]}$: 

\begin{equation}
\hat{C}_{yx}(f) = {1\over K} \sum_{k=1}^K \tilde y_k(f) \tilde x_k^*(f) 
\label{mtaper_corr}
\end{equation}

\begin{equation}
\hat{T}_{yx}(f) = { \sum_{k=1}^K \tilde y_k(f) \tilde x_k^*(f) 
\label{mtaper_trans} 
\over \sum_{k=1}^K \tilde x_k(f) \tilde x_k^*(f)} 
\end{equation}

\begin{equation}
\hat{\rho}_{yx}(f) =  {  \sum_{k=1}^K \tilde  y_k(f) \tilde  x_k^*(f) 
\over
\sqrt{ \sum_{k=1}^K |\tilde x_k(f)|^2 \sum_{k=1}^K |\tilde y_k(f)|^2}} 
\label{mtaper_coh}
\end{equation}

\setcounter{figure}{11} 
\begin{figure}
\centerline{\hbox{\psfig{figure=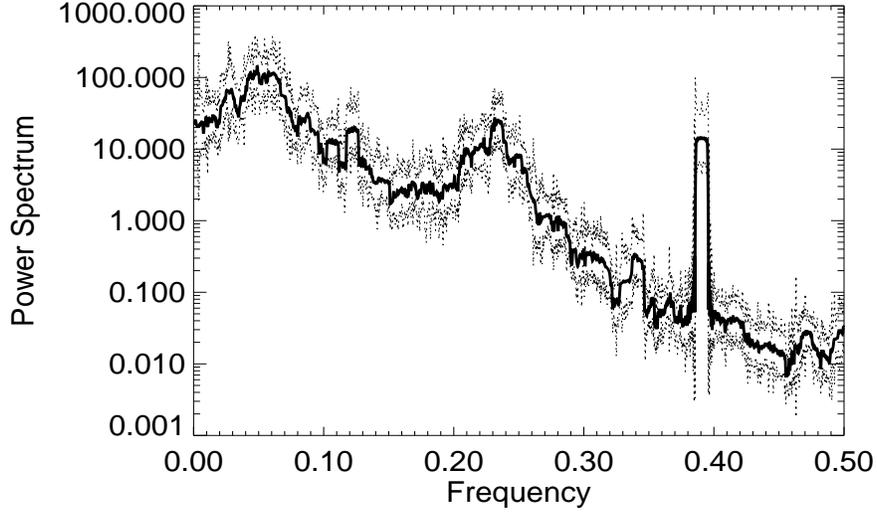,height=3in,width=4.5in}}}
\caption{
Jackknife error bars on multitaper spectral estimate
corresponding to spectrum shown in Fig. 6.  The spectral estimate is 
the solid bar and the error bars are represented by dots.
}
\end{figure}

These definitions allow the  estimation of the coherence and  transfer
function from a  single instance of  a pair of  time series. Using the
local frequency  ensemble, one can  also estimate jackknife error bars
for the spectra and the above quantities. The idea of
the jackknife is  to create different  estimates by leaving out a data
taper in turn. This creates a set of estimates from which an error bar
may be computed \cite{Thomson91}.  As an example, we show in Fig. 12
jackknife   estimates  of the   standard   deviations of  the spectral
estimate in Fig. 6. 

\section{Analysis techniques for different Modalities}

We now  deal  with  specific  examples of   data from different  brain
imaging modalities. We concentrate  on general strategies for  dealing
with  such data, and  most of our techniques  apply with minor changes
across the  modes. However, it  is easier  to treat  the various cases
seperately,  and  we present somewhat   parallel  developments in the
following sections. The data sets \A\ through \D\ have
been briefly discussed before. 

\subsection{Magnetoencephalography}

In this section, we consider data set \A\ consisting of multichannel MEG
recordings.  The number of channels is 74 (37 on each hemisphere), 
the digitization rate $2.083kHz$, and the duration of the recording is $5$ 
minutes. 

\subsubsection{Preprocessing}

An occasional artifact in MEG recordings  consists of regularly spaced
spikes in the  recordings due to cardiac activity.  This is  caused when 
the comparatively strong  magnetic field due to  currents in the heart
are not  well cancelled.  To suppress  this  artifact, we   proceed as
follows:   First, a space-time SVD is   performed on the multichannel
data. The  cardiac artifacts are  usually quite coherent in space, and
show  up in a  few  principal  component  time  series. In those  time
series, the  spikes  are segmented  out  by determining a threshhold  
by eye and segmenting out 62.5ms before and after the threshhold crossing.
The segmented heartbeat events are then averaged to determine a 
mean waveform.  Since the heartbeat amplitude is not constant across 
events
the heartbeat spikes are removed individually by fitting a scaling amplitude
to the mean waveform using a least-squares technique.  
Each spike modelled by the mean waveform multiplied by a scaling amplitude
 is then subtracted from the time series. Fig.13
illustrates the results of this procedure. 

\setcounter{figure}{12} 
\begin{figure}
\centerline{\hbox{\psfig{figure=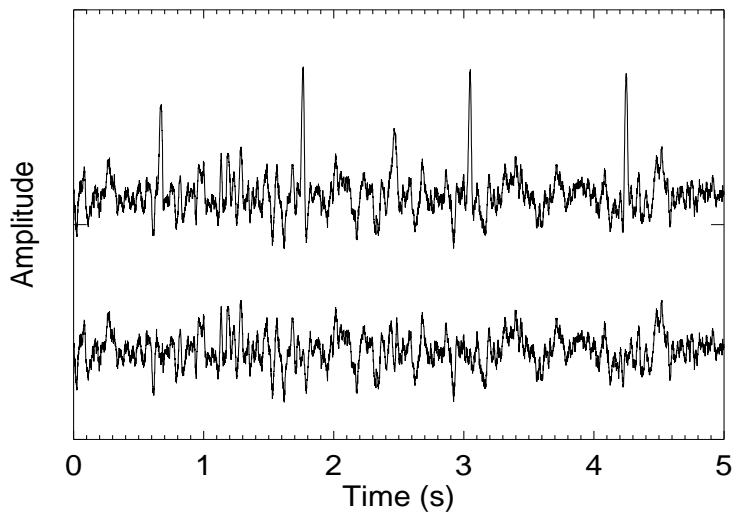,height=3in,width=4.5in}}}
\caption{
Principal component time series of MEG data from data set \A\
containing cardiac artifacts.  The upper plot shows the time course
before suppression and the spikes due to the heart beat are clearly
visible.  The lower plot shows the time course after suppression of the
heart beat using the technique described in the text.
}
\end{figure}

A fairly common  problem in electrical recordings  is the  presence of
$60Hz$ artifacts and on occasion sinusoidal artifacts at other
frequencies. Such sinusoidal artifacts,  if they lie in the  relevant
data    range, are usually dealt with    using notch  filters. This is
however unnecessarily  severe since the notch  filters  may remove too
large a band  of frequencies, in  particular in  the case of  MEG data
where frequencies  close to $60Hz$ are of  interest.  We find that the
$60Hz$ line  and other  fixed  sinusoidal artifacts can  be efficiently
estimated  and removed  using  the methods  for  sinusoidal estimation
described   in section 4.1.5.   Specifically, in this case the
frequency of the  line is accurately known (this requires a
precise  knowledge of  the  digitization rate)  and  one has to  only
estimate its  amplitude and  phase.  This is  done  for a  small  time
window  using  Eq.\ref{line_estimate}.  By  sliding this  time  window
along,  one obtains  a slowly  varying estimate  of  the amplitude and
phase,  and    is therefore  able  to  reconstruct   and subtract  the
sinusoidal artifact.  The results  of such a procedure are illustrated
in  Fig.14(a), where a  time averaged  spectrum is shown  for a single
channel   before   and after   subtraction     of the  line  frequency
component. In Fig.14(b), the amplitude of the  estimated 60Hz component 
is  shown with its slow modulation over time. 

\setcounter{figure}{13} 
\begin{figure}[t]
\begin{center}
\mbox{
\makebox[0 in][l]{$(a)$}\psfig{figure=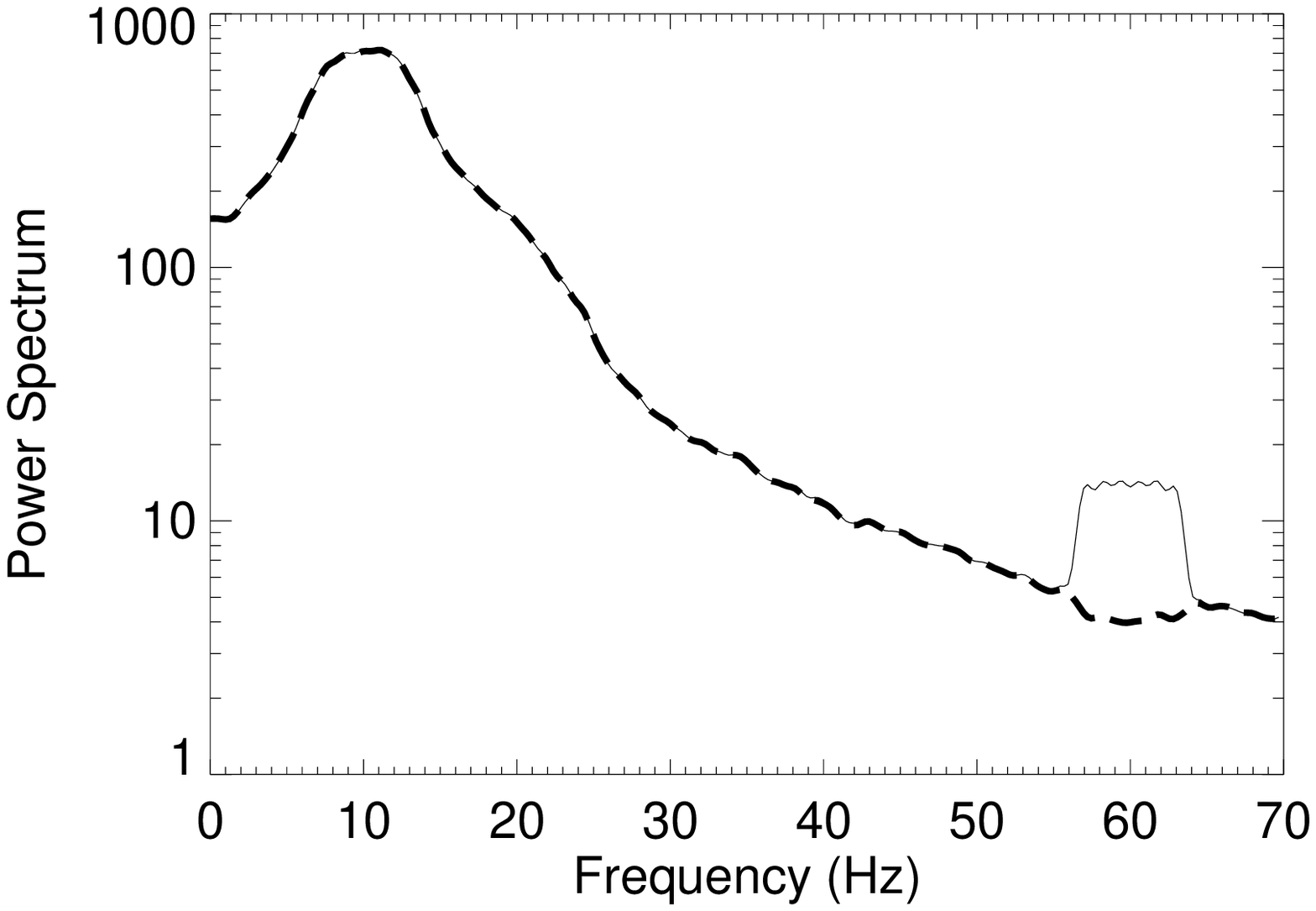,width=3.0 true
in,height=2.53 true in}
\hspace{0.1in} 
\makebox[0 in][l]{$(b)$}\psfig{figure=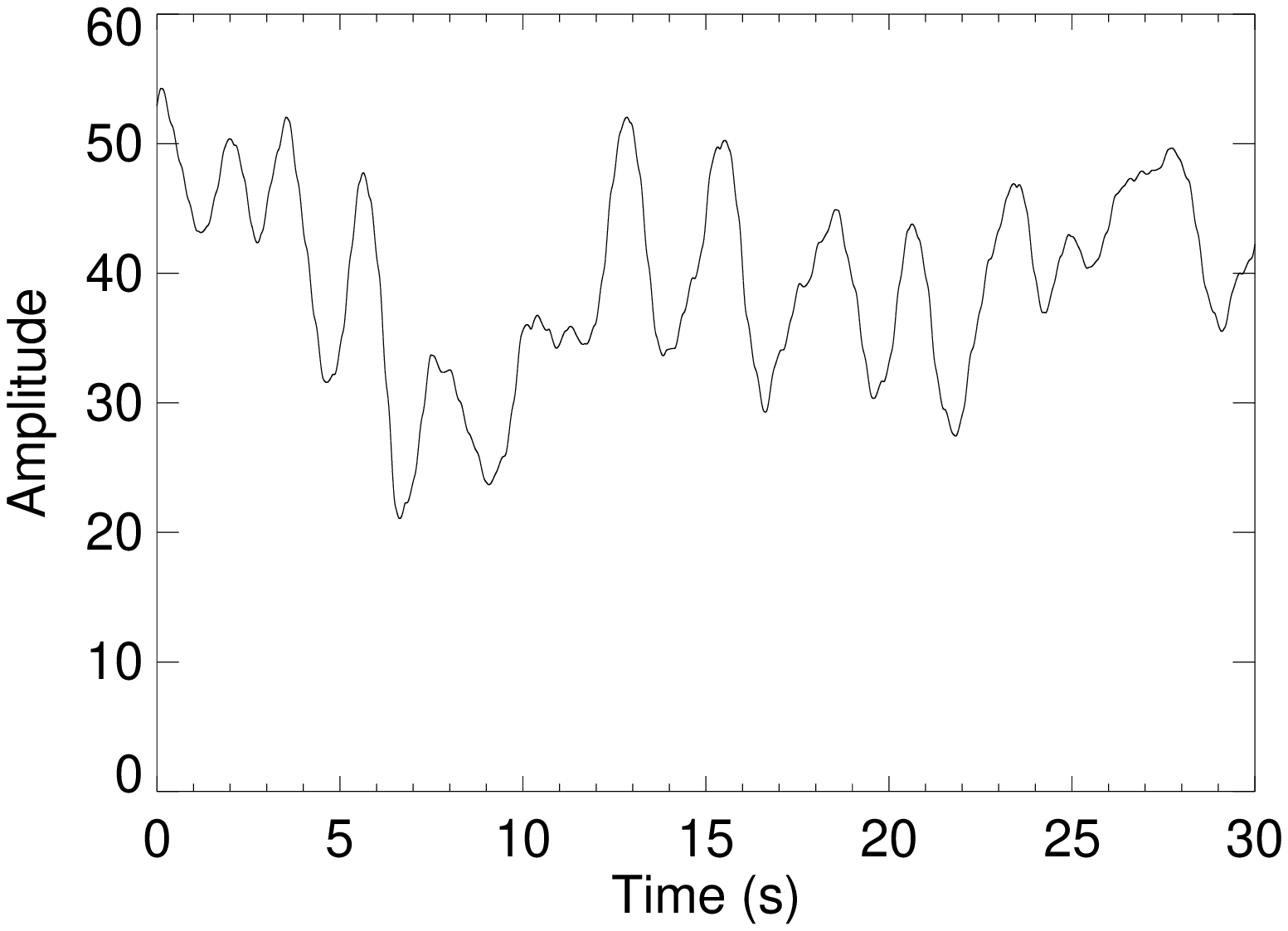,width=3.0 true
in,height=2.53 true in}   
}\\
\end{center}
\caption[abbrev]{ 
{\em a)} Time averaged spectrum before and after subtraction of 60Hz
artifact (single channel time series) from MEG data in data set \A . 
{\em b)} Estimated time varying amplitude of the 60 Hz line frequency. 
}
\end{figure}

\subsubsection{Time Frequency Analysis}

The   spectral analysis  of multichannel data   presents the fundamental
problem of how to simultaneously visualize or otherwise examine
the time-frequency content  of many channels.   One way  to reduce the
dimensionality of the problem is to work with principal component time
series. A space-time SVD of MEG data gives a rapidly decaying singular
value spectrum. This indicates  that one  can  consider only the  first few
temporal   components  to  understand  the   spectral content   of the
data. For purposes of  illustration of our techniques we consider
the first three principal component time  series in a 5 minute segment
recording of spontaneous awake activity. 

\setcounter{figure}{14} 
\begin{figure}
\centerline{\hbox{\psfig{figure=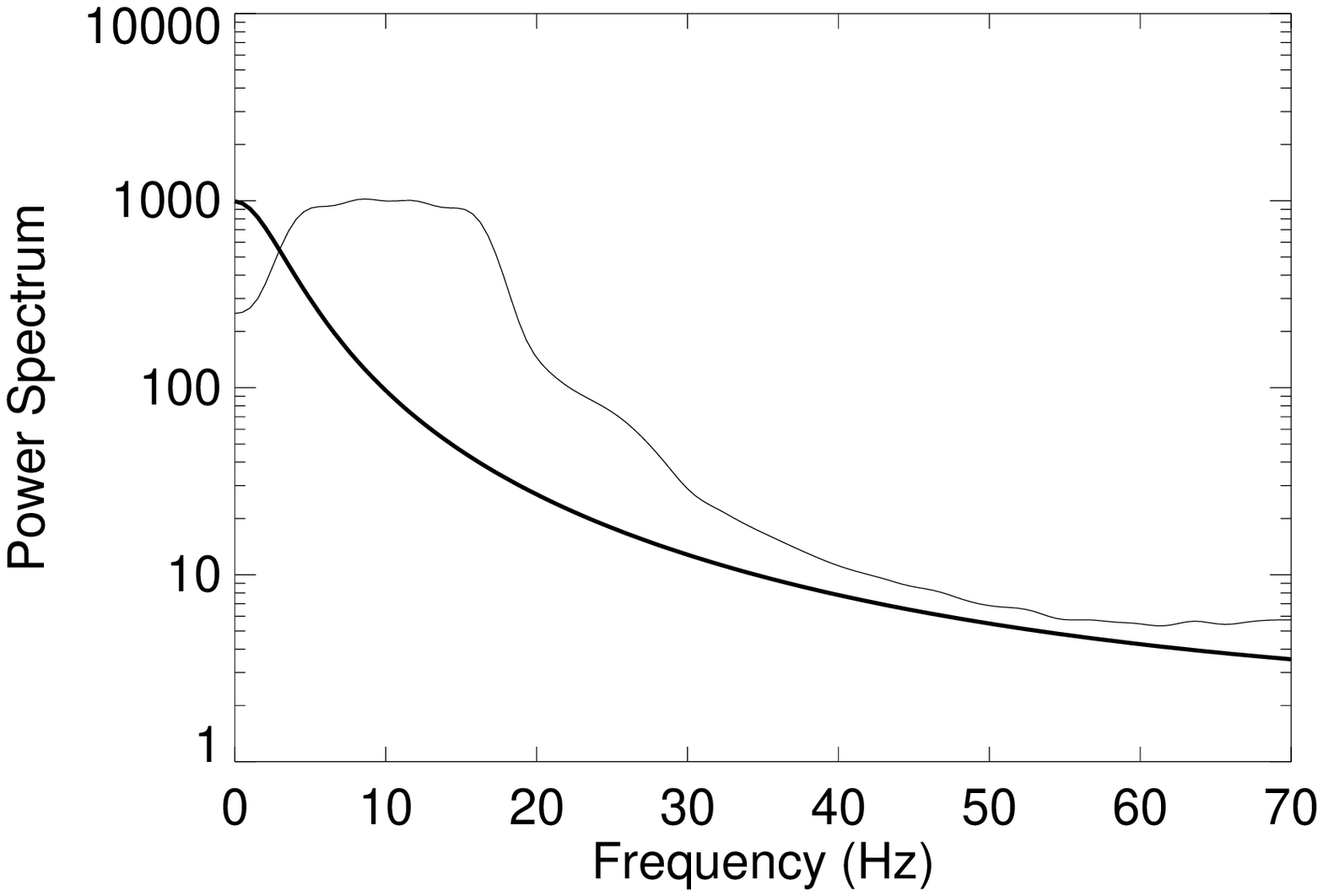,height=3in,width=4.5in}}}
\caption{Autoregressive fit to averaged spectrum of MEG data from data
set \A\ averaged over non-overlapping 
windows in time for purposes of pre-whitening.  The thick line shows
the autoregressive fit to the average spectrum. 
The thin line shows the average spectrum.  A low-order AR spectral 
estimate is used to reduce the dynamic range in the spectrum without
fitting specific structural features of the data. 
}
\end{figure}

A useful preliminary step in spectral  analysis  
  is pre-whitening  using  an appropriate autoregressive model.
This   is  necessary because  the   spectrum   has a large   dynamic
range.   Pre-whitening   leads   to   equalization   of   power across
frequencies,   which  allows  better  visualization of  time-frequency
spectra.   In addition,  a  space-time SVD   is   better  performed on
pre-whitened data, since  otherwise the large amplitude, low frequency
oscillations  completely  dominate  the  principal    components.  The
qualitative   character of the   average  spectrum  is   a slope  on a
semilogarithmic scale (Fig.14(a)). The goal is to prewhiten with a low
order AR model so that the peaks in the spectrum are left in place but
the  overall  slope  is removed.  Considering  the derivative  of  the
spectrum rather than the spectrum itself achieves a similar result. 
 
The procedure  is to first calculate a moving  estimate of  the spectrum 
using  a short time window ($T  = 0.35s$  in the  present  case) and a
direct multitaper  spectral  estimator ($ W  = 4/T;  K  = 6 $).  These
estimates  are then  averaged over  time  to obtain  a  smooth overall
spectrum. Next, a low order autoregressive model (order $ = 10$ in the
present case) is fit to the spectral estimate. We use the Levinson Durbin recursion
to fit the AR  model. Results of such a  fit are shown  in Fig.15.  The
coefficients of the autoregressive process are then used
to   filter the data,   and the residuals   are subjected  to  further
analysis.  Thus,  if  the  coefficients obtained   are $a_k$  and  the
original time  series is $x_t$, then  the residuals are  $\delta x_t =  x_t -
\sum_1^n x_{t-k} a_k$ which are subjected to a time-frequency analysis. 

Typical time frequency spectra of pre-whitened principal component time
series are  shown in Fig.16. The spectra  were obtained using a direct
multitaper estimate for $1 s$ long time windows and  for $W = 4Hz, K =
6$. 

\setcounter{figure}{15}
\begin{figure}
\centerline{\hbox{\psfig{figure=figure16.ps,height=6in,width=6in}}}
\caption{
Time frequency spectrum of first three principal components
obtained in a space-time SVD of MEG data from data set \A\ pre-whitened by 
the filter shown in Figure 15.
}
\end{figure}

\setcounter{figure}{16} 
\begin{figure}[t]
\begin{center}
\mbox{
\makebox[0 in][l]{$(a)$}\psfig{figure=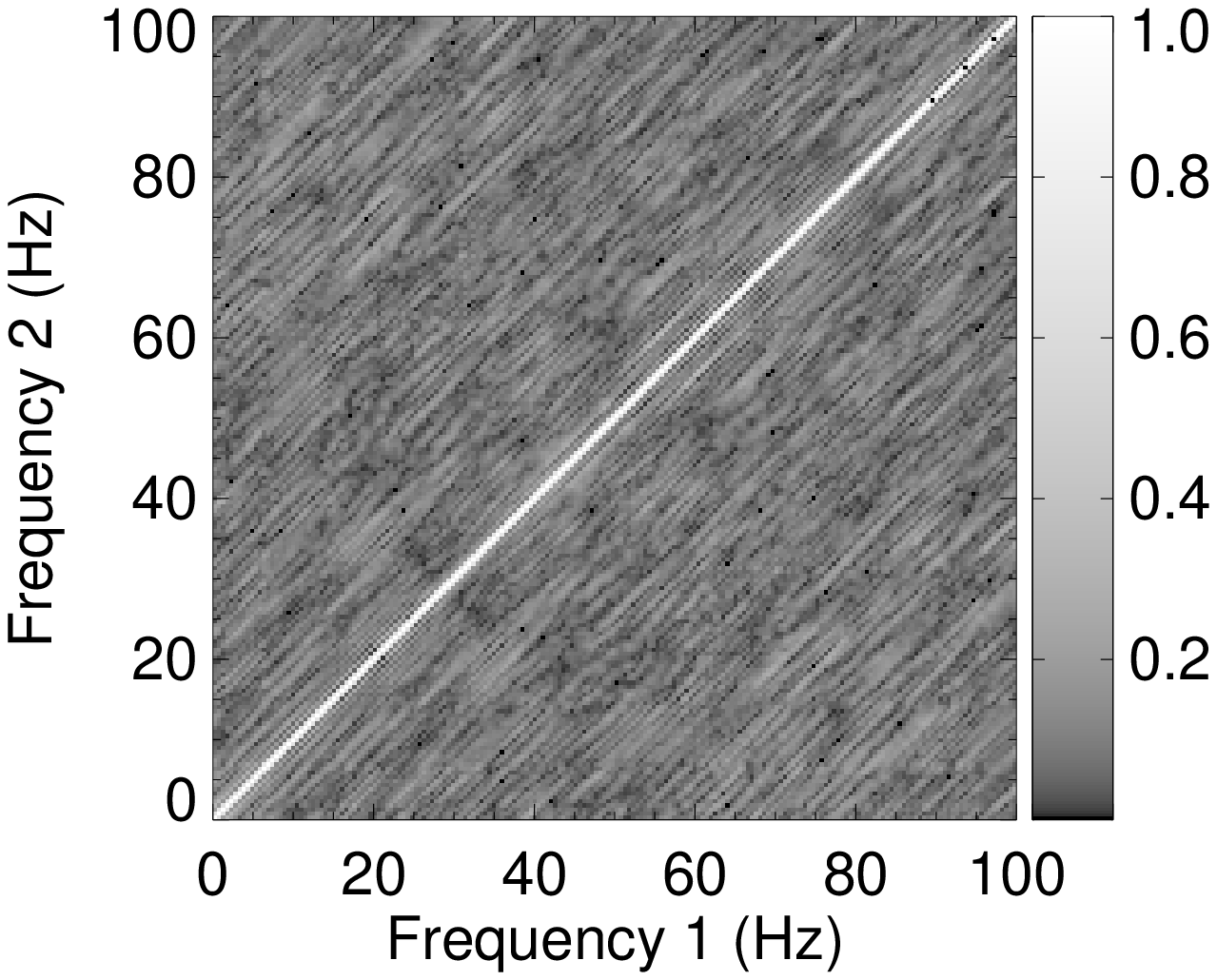,width=3.0 true
in,height=2.53 true in}
\hspace{0.1in} 
\makebox[0 in][l]{$(b)$}\psfig{figure=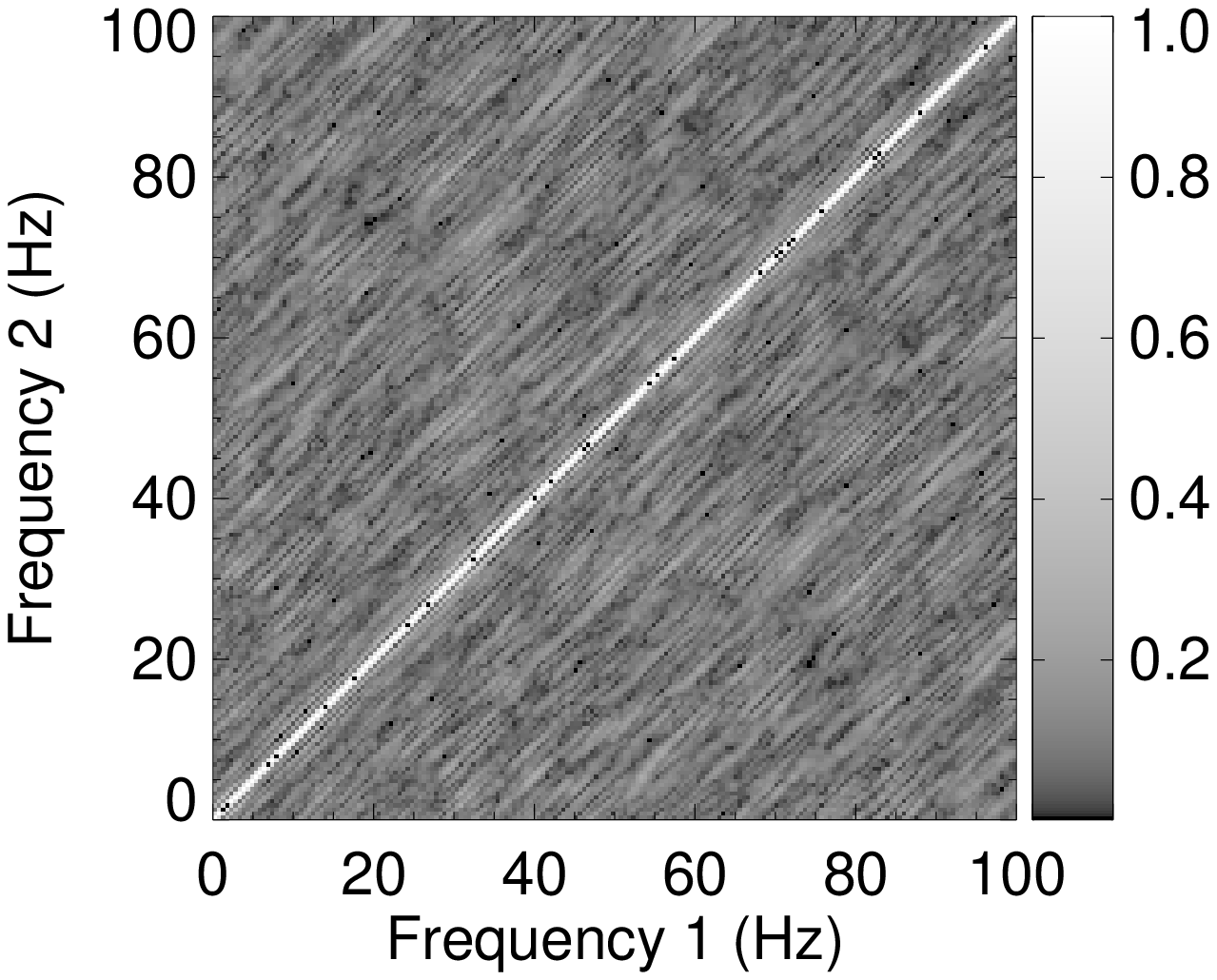,width=3.0 true
in,height=2.53 true in}   
}\\
\end{center}
\caption[abbrev]{ 
{\em a)} Magnitude of $\rho(f,f^{\prime})$ for the leading PC time
series of MEG data from data set \A .
{\em b)} Magnitude of $\rho(f,f^{\prime})$ for the leading PC time
series after initially scrambling the time series.
}
\end{figure}

To assess the quality of the spectral characterization it is important
to quantify   the  presence (or  absence)   of  correlations between
fluctuations at different frequencies. One measure of the correlations
between frequencies for a given time  series is given by the following
quantity analogous to the coherence between different channels:
 
\begin{equation}
\rho_{xx}(f,f^{\prime})   =  {  \sum_{k=1}^K    \tilde  x_k(f)  \tilde
x_k^*(f^{\prime})   \over  \sqrt{  \sum_{k=1}^K  |\tilde      x_k(f)|^2
\sum_{k=1}^K |\tilde x_k(f^{\prime})|^2}} 
\label{mtaper_coherence}
\end{equation}

The estimate  can be further  averaged across time windows to increase
the  number  of degrees of  freedom.   This quantity,
computed  for the  leading  principal component  time series  obtained
earlier,  is displayed in Fig.17(a). In  this figure, the magnitude of
the estimate $\rho(f,f^{\prime})$  for  the leading PC time  series is
displayed as a   function of the   two arguments, $f$ and $f'$.  
For comparison,  in
Fig.17(b),  results  of the  same  procedure  obtained after initially
scrambling the time series  are  also displayed. A visual   comparison
shows the lack of evidence for correlations  across frequencies in an
average sense. 

\setcounter{figure}{17} 
\begin{figure}
\centerline{\hbox{\psfig{figure=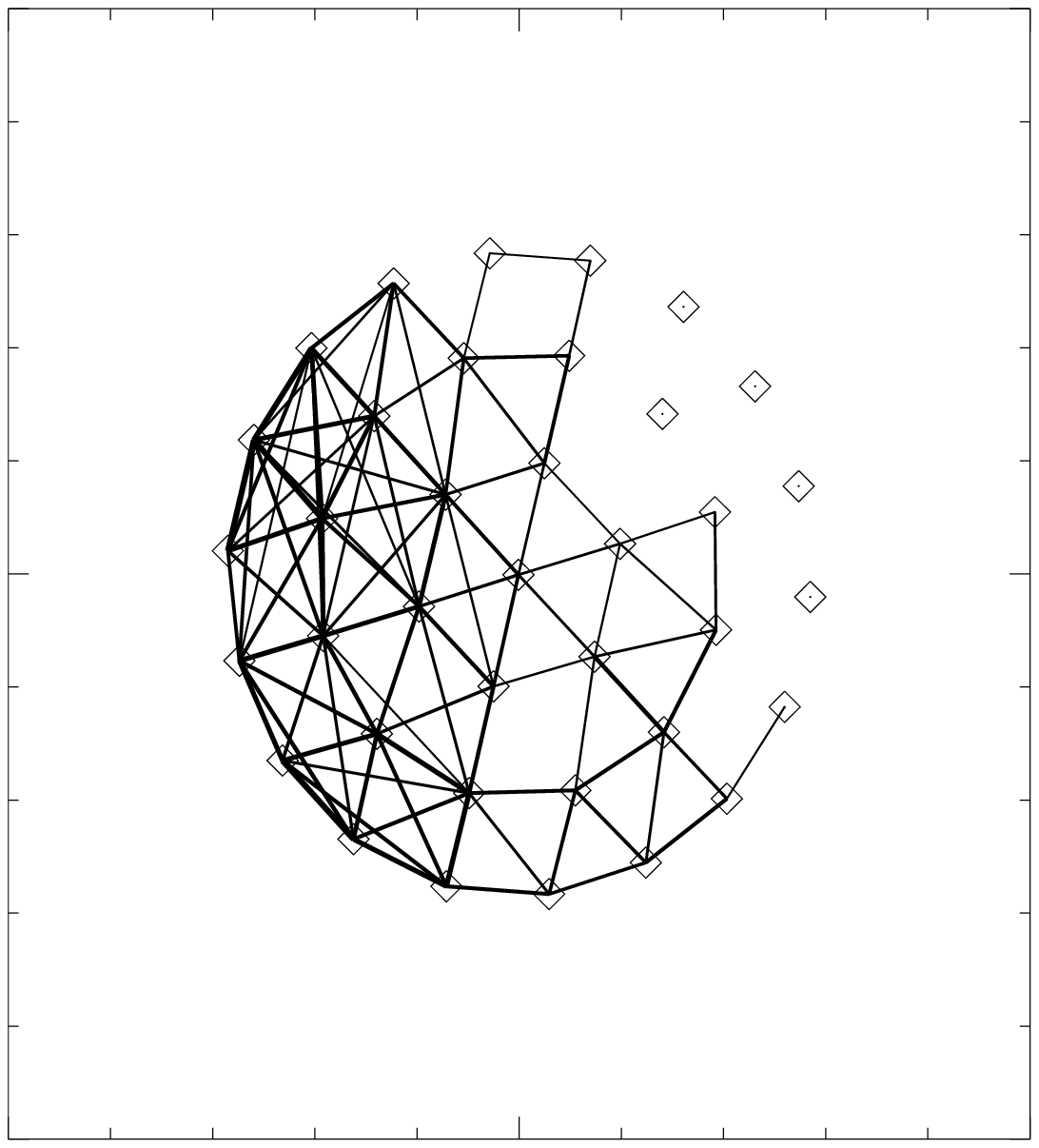,height=3in,width=3in}}}
\caption{
Magnitudes of coherences for MEG from data set \A\ 
between points in space displayed by lines
of proportionate thickness. The coherences were computed for a center
frequency of $20Hz$ and half bandwidth $5Hz$.
}
\end{figure}

\subsubsection{Multichannel Spectral Analysis}

The time frequency  spectra of leading  principal components  shown in
the earlier section capture the spectral content of the MEG signal
that is coherent in space. Alternatively, one
can perform a space-frequency SVD with a moving time window on the data. 
It is also desirable to obtain time averaged
characterizations of the coherence across channels.   This can be done
by  considering     the  coherence     functions   between    channels
$\rho_{ij}(f)$, which  can  be estimated   as in  Eq.\ref{mtaper_coh}
using multitaper methods. The estimate, when calculated for a
moving time  window,     can   be   further    averaged    across   time
windows. Displaying   the  matrix $\rho_{ij}$ poses  a  visualization
problem, since the indices $i,j$ themselves correspond to locations on
a two dimensional grid.  Thus, an image displaying the matrix
$\rho_{ij}$  does  not  preserve  the spatial   relationships between
channels. One  solution to this  visualization problem is presented in
Fig.18, by representing  the   strength of the coherence  between  two
space  points by  the  thickness    of  a bond  connecting  the    two
points. This figure shows  the coherence $\rho_{ij}(f)$ computed with
a center frequency $20Hz$ and half bandwidth $5Hz$. The bond strengths
have been threshholded to facilitate  the display. This visualization,
although   not   quantitative, allows    for an   assessment    of the
organization of the coherences in space. 

\setcounter{figure}{18} 
\begin{figure}
\centerline{\hbox{\psfig{figure=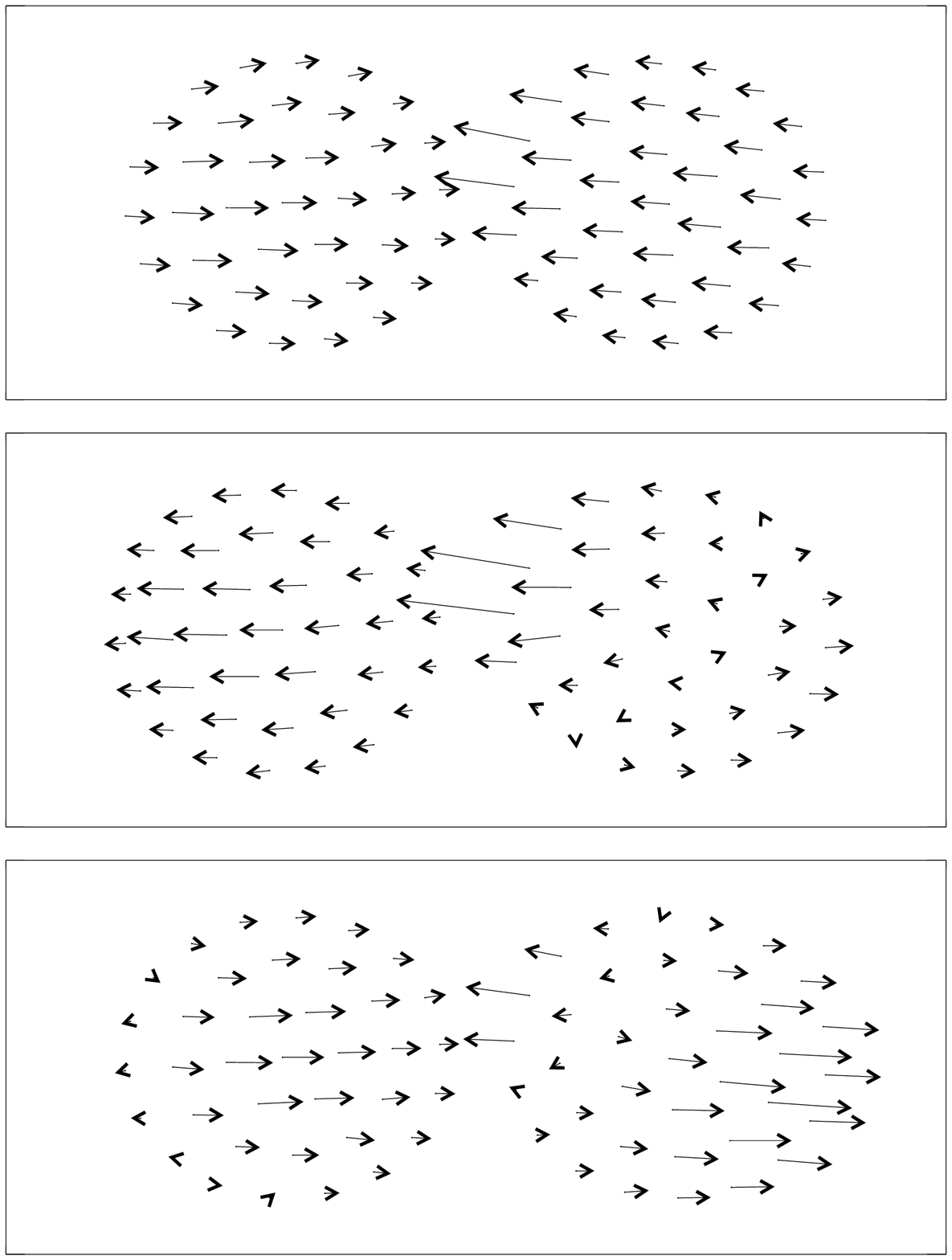,height=6.in,width=6in}}}
\caption{
Dominant spatial eigenmodes of a space-time SVD of band pass
filtered MEG data from data set \A\ for the frequency band 35-45 Hz. 
The two hemispheres of the 
brain are projected onto a plane, each being sampled by 37 sensors.
}
\end{figure}

An alternative way   of  performing principal   component analysis  on
space-time  data while localizing information  in the frequency domain
is clearly to apply  a  space-time SVD to  data  which has first  been
frequency filtered into the  desired band. To obtain frequency filters
with optimal bandlimiting properties, projection filters based on DPSS
are  used. For illustration,  we  perform this analysis on
the data under discussion. The individual channels were first filtered
into the frequency band  $35-45 Hz$. This gives  a complex time series
at each spatial location. The  first three dominant spatial eigenmodes
in a space-frequency SVD are displayed in Fig. 19, with singular values
decreasing from the top to the bottom of the figure. Since the spatial
eigenmodes are  complex, their values are  represented by 
arrows, whose  lengths correspond to  magnitudes and whose directions 
correspond to  phases.  It is quite clear  from the figure that
the data shows a high degree of spatial coherence on an average. 

\subsection{Optical Imaging}

In this  subsection we consider  optical imaging data. We consider the
general  case of  imaging  data  gathered  either  using intrinsic  or
extrinsic contrast.  The case of intrinsic contrast is closely related
to fMRI, and the analysis parallels that of the fMRI data sets \C\ and
\D. 
To illustrate   some  effects  important in   the   case of  extrinsic
contrast, we consider data set \B.  Data set \B\ was gathered in
presence
of a  voltage sensitive  dye, and consists   of images of   a isolated
pro-cerebral  lobe of Limax,  with a  digitization   rate $75 Hz$  and
duration $23 s$. 

\subsubsection{Spectral Analysis of PC Time Series}

The general procedure  outlined  above for  fMRI data consisting  of a
space-time SVD followed  by  a  spectral  analysis of  the   principal
component (PC)    time series  is     useful   to  obtain   a 
preliminary
characterization of the data. In   particular, in the case of  optical
imaging data using intrinsic or extrinsic  contrast in the presence of
respiratory and cardiac artifacts, this procedure helps the assessment
of the artifactual content of the data. However, as discussed earlier,
the space-time SVD mixes  up distinct dynamic  components of the image
data, and  is therefore of limited  utility for a full characterization of
the data. 

\subsubsection{Removal of Physiological Artifacts}

We have  developed a method for  efficient suppression  of respiratory
and cardiac  artifacts  from brain  imaging   data (including  optical
images and MRI images for high digitization  rates) by modelling these
processes by slowly    amplitude and frequency  modulated   sinusoids.
Other approaches to suppression  of these artifacts in  the literature
include `gating',  frequency filtering, modelling  of the
oscillations  by a periodic function \cite{Le96},  and removal of
selected components  in  a space-time SVD \cite{Orbach95}.   These
approaches have varying efficacies. For example, gating aliases the relevant 
oscillations down
to  zero frequency, so that  any variation in these oscillations cause
slow   fluctuations    in    the   data,   which    is   in    general
undesirable. Frequency filtering removes  more spectral energy than is
strictly necessary from the signal. Modelling of the oscillations by a
periodic function is imperfect because the oscillations themselves may
vary in time. This can be rectified by allowing  the parameters of the
oscillations to  slowly  change in time.   One must be  able to  fit a
sinusoidal model robustly  to  short time series  segments to  do this
properly.  Finally,  removing selected components  in a space-time SVD
is not  a safe procedure,  since, as discussed before,  the space-time
SVD  does  not necessarily  separate  the different  components of the
image data. 

Our method for  suppressing the above mentioned oscillatory components
is based on multitaper methods  for estimating sinusoids in a  colored
background  described earlier.  The method  is  based on modelling the
oscillations by a  sum of sinusoids whose  amplitude and frequency are
allowed to vary slowly. The modeled oscillations are removed from the
time  series in the data to  obtain  the desired residuals. Although a
space-time SVD  is not sufficient  by itself, applying such  a removal
technique    to   the  leading  temporal    principal components,  and
reconstituting     the residual  time   series   appears  to give good
results. This procedure  was found to  be effective for a wide variety
of data, including optical imaging  using both intrinsic and extrinsic
contrast in rat brain, and in fMRI data.  The details of the technique
are presented in the section on fMRI data, section 5.3.

\subsubsection{Space-Frequency SVD}

\setcounter{figure}{19} 
\begin{figure}
\centerline{\hbox{\psfig{figure=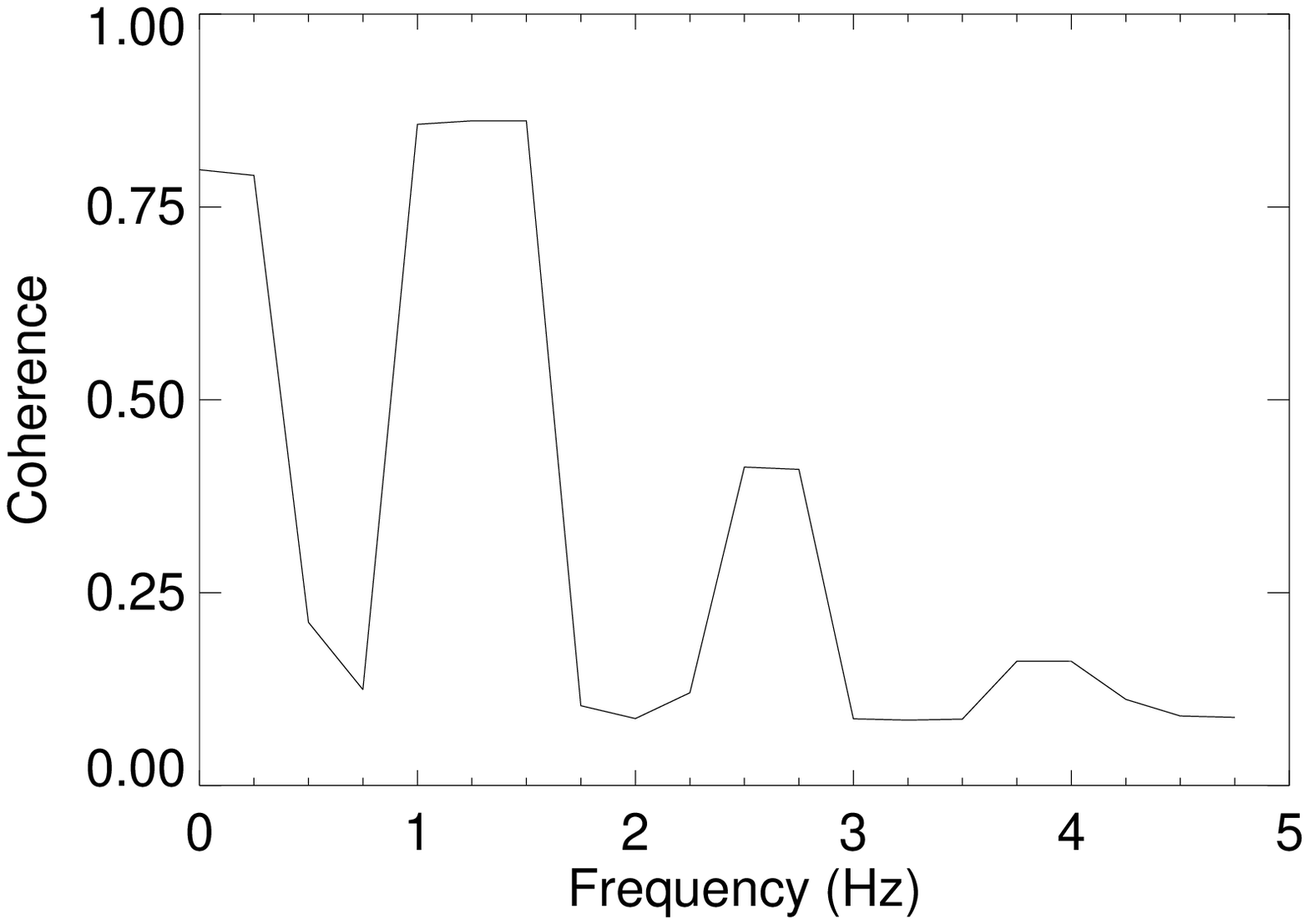,height=3in,width=4.5in}}}
\caption{
Coherence spectrum for a space-frequency SVD of 
optical imaging data in data set \B\ from the PC lobe of Limax. 
}
\end{figure} 

In  this section   we present  the results  of  a space-frequency  SVD
applied to   data  set \B.  This  technique  has  been  described above
(Fig.10 and 11) for  data set  \D.   In Fig  20, the coherence  spectrum is
shown for data set  \B. The coherence   was computed on a coarse  grid,
since there  is not  much finer structure  than  that displayed in the
spectrum. In  this  case, 13 DPSS  were used,  corresponding to a full
bandwidth of $0.3Hz$.  The  preparation (pro-cerebral lobe of Limax) is
known to   show oscillations,  which  are organized   in  space as   a
traveling  wave.  Traveling  waves   in the image,   gathered in the
presence of a  voltage sensitive dye, reflect  traveling waves in the
electrical activity.  The coherence spectrum displayed in Fig.20 shows
a fundamental  frequency of about  1.25Hz  and the corresponding first
two harmonics. 

The amplitudes  of the  leading  spatial eigenmode  as a   function of
center   frequency  are   shown  in Fig.21.   Note   that the  spatial
distribution of coherence is more localized to the center of the image
at the higher  harmonics.  This reflects the  change  in shape  of the
waveform of oscillation that is known to occur in the preparation as a
function of spatial position.  This phenomenon has been
interpreted as a result of differing spatial concentration profiles of
two  different cell types  in this  system \cite{Kleinfeld94}. Based  on
this   past interpretation, the leading  modes  at the fundamental and
harmonic frequencies directly reflect  the spatial distribution of the
different cell types. 

\setcounter{figure}{20} 
\begin{figure}
\centerline{\hbox{\psfig{figure=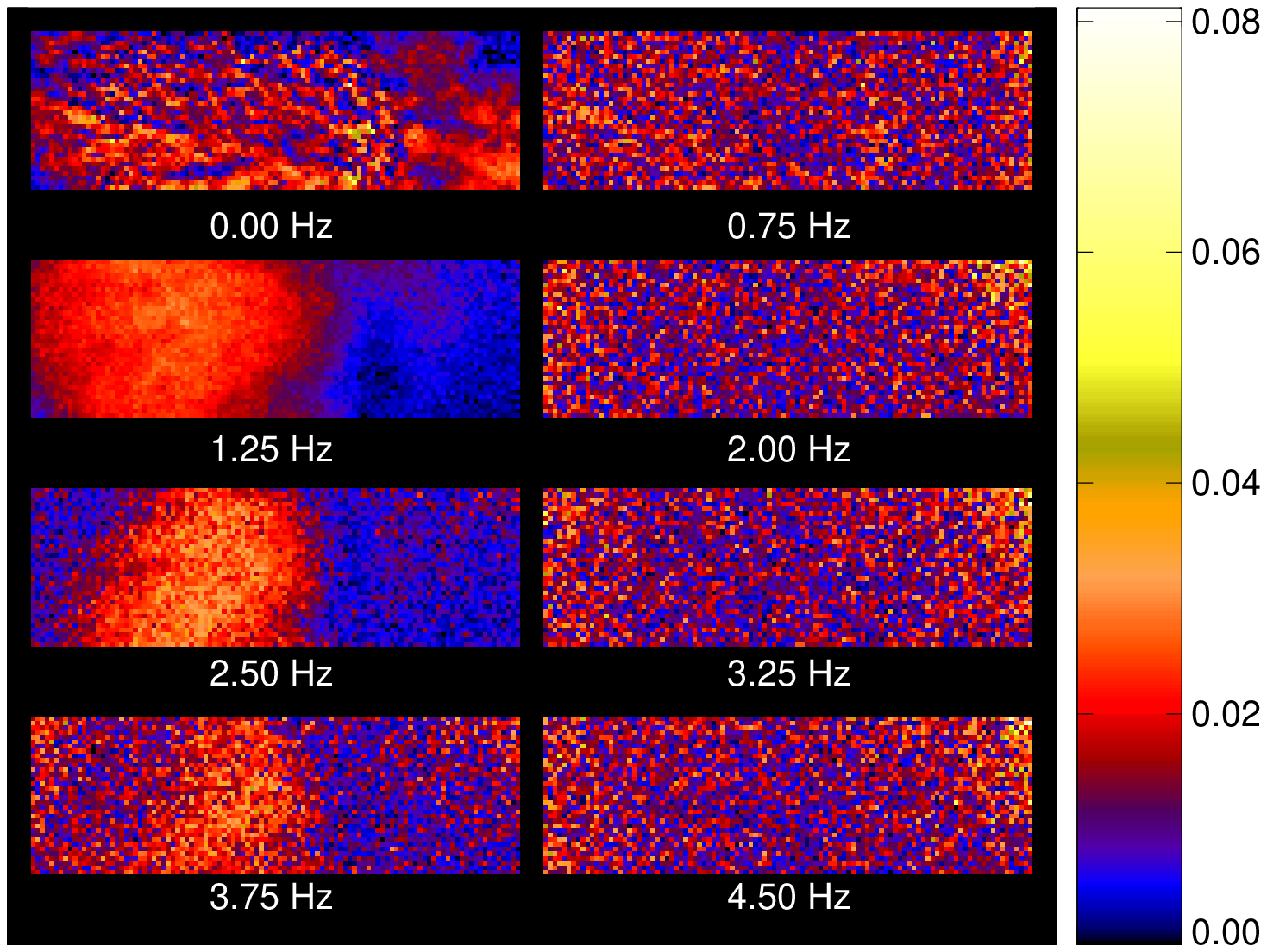,height=3in,width=4.5in}}}
\caption{
Leading spatial eigenmodes of the space-frequency SVD
corresponding to Fig. 20.  The amplitudes of the leading modes are 
shown as a function of center frequency.
}
\end{figure}

\setcounter{figure}{21} 
\begin{figure}[t]
\begin{center}
\mbox{
\makebox[0 in][l]{$(a)$}\psfig{figure=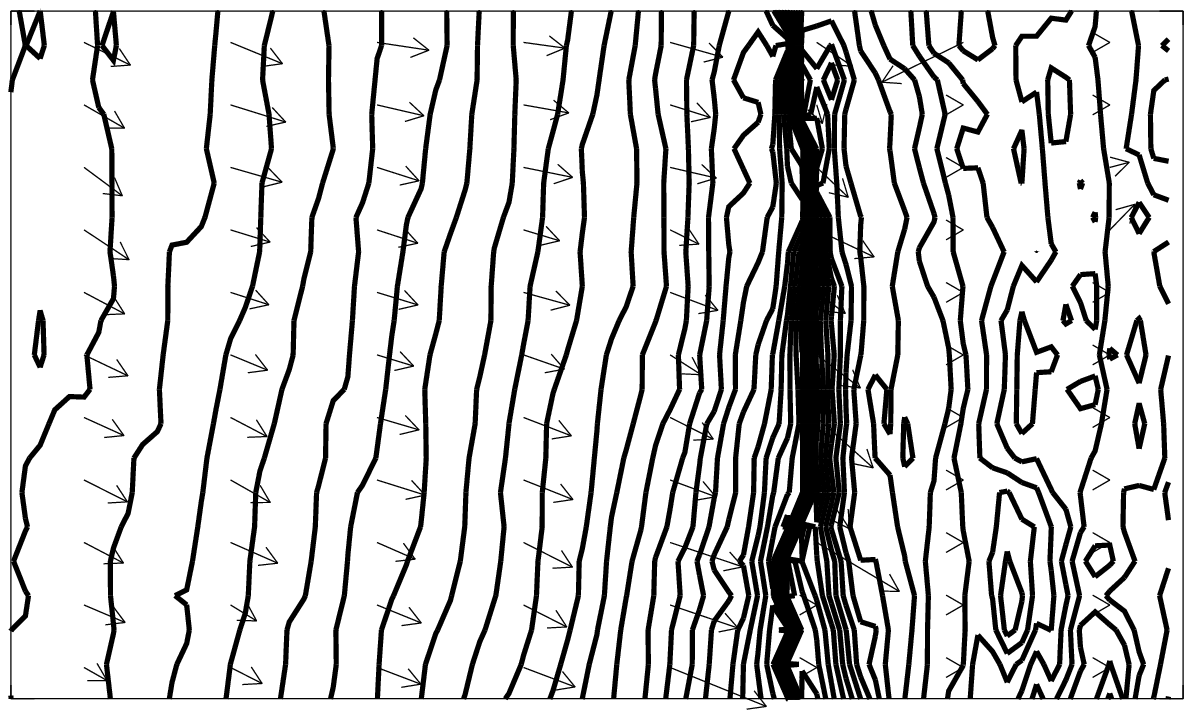,width=3.0 true
in,height=2.53 true in}
\hspace{0.1in} 
\makebox[0 in][l]{$(b)$}\psfig{figure=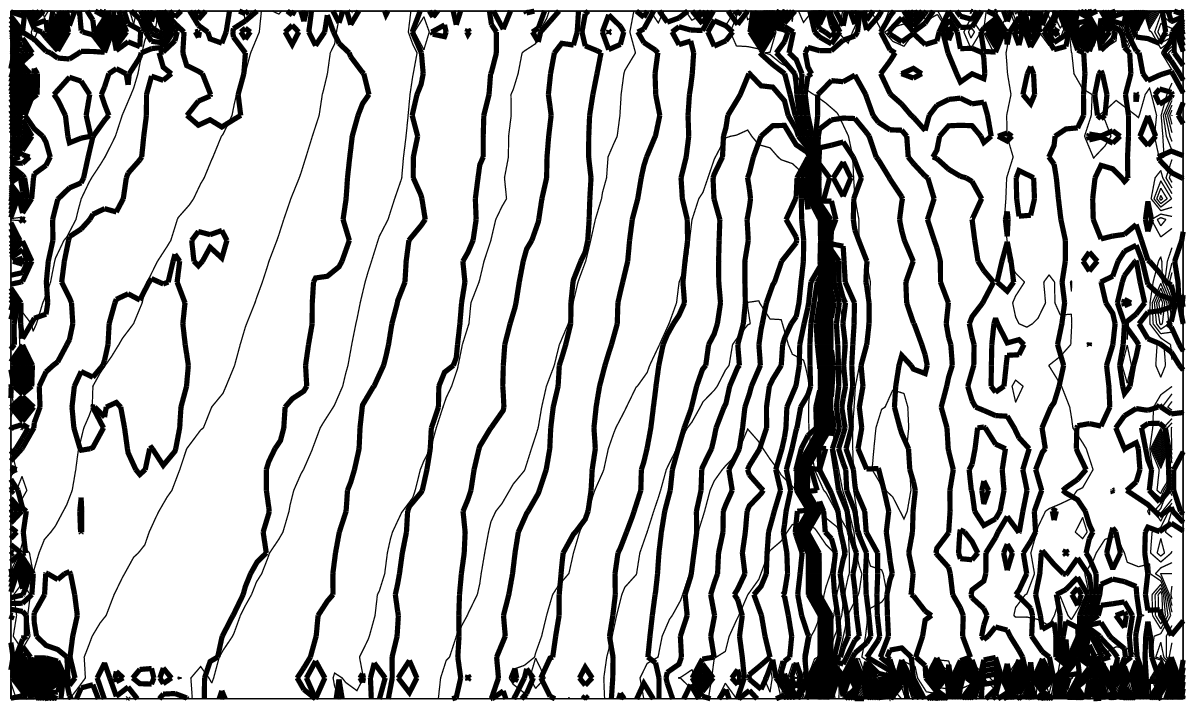,width=3.0 true
in,height=2.53 true in}   
}\\
\end{center}
\caption[abbrev]{
{Leading spatial eigenmodes of the space-frequency SVD
corresponding to Fig.21.}  
{\em a)} Gradients of the phase of the leading SVD mode at a center
frequency of 2.5Hz,  corresponding to local 
wave-vectors for the wave motion, are shown as arrows. These are 
superposed on constant phase contours for the mode.
{\em b)} Constant phase contours for spatial
eigenmodes of the space-frequency SVD for center 
frequencies 1.25 Hz and 2.5 Hz.  2.5 Hz is shown by the bold contour. 
}
\end{figure}

The spatial eigenmodes are complex, and possess a phase in addition to
an  amplitude.  This allows for  the investigation of traveling waves
in the data.  Let  the  leading spatial  eigenmode,  as a function  of
frequency, be expressed as 

\begin{equation}
\tilde I_1(x;f) = A(x;f)e^{i \theta(x;f)} 
\end{equation}

Given   the  convention we are   following   for the Fourier transform
(Eq.\ref{fft_convention}),  one may define   the following  local wave
vector: 

\begin{equation}
{\bf k}(x;f) = -\mbox{\boldmath $\nabla $} \theta(x;f) 
\end{equation}

If   the coherent fluctuations    at a given   frequency correspond  to
traveling plane waves, then  ${\bf k}(x;f)$ corresponds to  the usual
definition of the wave  vector. More  generally, this quantity  allows
the systematic examination   of phase gradients  in  the system, which
corresponds to traveling excitations. 

The local wavevector map  for data set \B\ at  a center frequency of 2.5
Hz (the  first harmonic)  is shown in   Fig.22(a),  superposed on top of
contours of constant phase.  In Fig.22(b), the constant phase contours
for center frequencies 1.25Hz, 2.5Hz  (fundamental and first harmonic)
are  superposed.  On superposing  the contours for the fundamental and
the first harmonic, we discover an effect  that was not evident in the
earlier analysis of the  data \cite{Kleinfeld94}, namely that the  phase
gradients at 1.25Hz and 2.5Hz are slightly tilted with respect to each
other.  This can be  interpreted as two different waves simultaneously
present in the  system, but running  in slightly different directions.
Coexisting   waves  present at  different   temporal frequencies  with
different  directions of  propagation   have  also been  revealed   by
space-frequency SVD analysis of voltage-sensitive dye images of turtle
visual cortex \cite{Prechtl97}. 

\subsection{Magnetic Resonance Imaging}

The data sets \C\ and \D\ comprising functional MRI data have been used
in
earlier  sections to illustrate the techniques  presented in the paper
(Figures 8-11). In this section,  we continue to illustrate analytical
techniques on this  data set.  The data sets were gathered
with a digitization rate of $5Hz$ and a total duration of $110s$. Data
set \C/\D\ was gathered in the presence/absence of a flashing LED
checkerboard pattern serving as visual stimulus.   An extra problem in
the analysis of MRI data is the presence  of motion related artifacts,
which have to be suppressed \cite{Mitra97}. 

\subsubsection{Removal of Physiological Artifacts}

Here a detailed description is  provided of the  method for removal of
physiological oscillations discussed in the section on optical imaging
data.  A space-time  SVD of the data  is  first computed, followed  by
sinusoidal  modelling   of  the  leading   principal  component   time
series. This is necessary for two  reasons: (i) The images in question
typically have  many pixels,  and it  is   impractical to perform  the
analysis separately on all pixels.  (ii) The leading SVD modes capture
a large degree of global coherence in the oscillations. 

Consider a single principal  component time series, $a(t)$.   We assume
that the time series is a  sum of two  components. The first component
consists  of  a sum  of  amplitude  and frequency  modulated sinusoids
representing  respiratory   and  cardiac   oscillations.    The second
component $\delta a(t) $ contains the desired signal. 
 
\begin{equation}
a(t) =  \sum_n A_n(t) \cos\bigl(f_n(t)  t + \phi_n(t)  \bigr) + \delta
a(t) 
\label{art_eqn}
\end{equation}

The goal  is to estimate  the  smooth functions $A_n(t), f_n(t)  $ and
$\phi_n(t)$,  which  give the   component  to be  subtracted  from the
original time series. 

It is necessary  to choose an  optimally  sized analysis window.  This
window must be  sufficiently  small to capture  the  variations in the
amplitude,  frequency and phase, but must  be long  enough to have the
frequency resolution  to separate the  relevant peaks in the spectrum,
both artifactual and originating in the desired  signal. The choice of
window  size depends to  some extent  on  the nature  of the data, and
cannot be easily automated.  However, in similar experiments it is safe
to use the same parameters. Ideally, one  would choose the window size
in some adaptive   manner, but we find  it  adequate  for  our present
purposes to work with a fixed window size. 

The frequencies $f_n(t)$ have to satisfy several criteria. Usually one
is removing the respiratory and  cardiac components. The corresponding
spectra contain   small    integer   multiples  of two     fundamental
frequencies, with the possible presence  of sidebands due to nonlinear
interactions between the oscillations.  The frequency F-test described
in section 4.1.5 is used to determine the fundamental frequency tracks
$f_n(t)$  in Eq.\ref{art_eqn}. The time  series used for this purpose
may either be  a principal component time  series, or an independently
monitored physiological time series. Note  that the assumption here is
that within an analysis bandwidth of the relevant  peak in the cardiac
or respiratory cycle,  the data can be  modeled  as a sine wave  in a
locally white background.  The fundamental frequency tracks are used to
construct the tracks for the harmonics and the sidebands.

\setcounter{figure}{22}
\begin{figure}
\centerline{\hbox{\psfig{figure=figure23.ps,height=3in,width=4.5in}}}
\caption{
Time frequency plot of frequency tracks for a moving 
sinusoidal model of the cardiac and respiratory artifacts in 
fMRI data from data set \C . The analysis is performed on 
a principal component time series obtained by performing a space-time
SVD.
}
\end{figure}

\setcounter{figure}{23} 
\begin{figure}
\centerline{\hbox{\psfig{figure=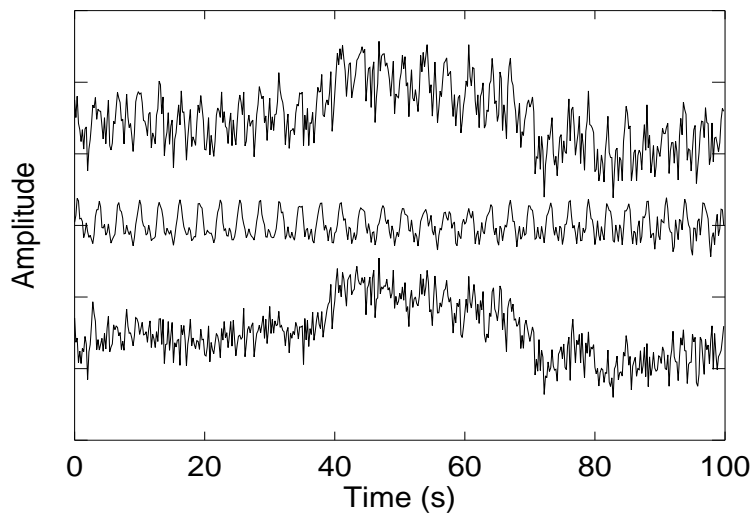,height=3in,width=4.5in}}}
\caption{Upper:  Part of principal component time series forming the
basis of Fig.23. 
Middle: Estimated cardiac and respiratory components corresponding to 
Upper plot.
Lower: Time series with cardiac and respiratory components suppressed.
}
\end{figure}

In the example, we  show the results of the  analysis on one principal
component time series  from   data set \C.  The  fundamental  frequency
tracks were determined by using the F-test on a moving analysis window
on the PC  time  series.  

If the F-test does not provide a frequency estimate for short segments
of the   data,  estimates may be  interpolated   using  a  spline, for
example.  After the frequency tracks are determined, the amplitude and
phase  of the  sinusoids  are calculated using Eq.\ref{line_estimate}
for each analysis window location. Note that the shift in time between
two successive analysis  windows can be  as small as the  digitization
rate of  the  data,  but  is limited   in  practice by  the  available
computational resources.  The  estimated  sinusoids are  reconstructed
for    each  analysis  window,  and   the   successive estimates   are
overlap-added to provide the final model waveform for the artifacts.

The frequency  tracks  are shown in   Fig.23
superposed on   a  time-frequency spectral  estimate of  the principal
component time series. 
In Fig.24, results of the procedure described  above are shown in the
time domain for the chosen principal component.  
Notice the change in the   frequency
corresponding to the cardiac cycle (around 1.3Hz)  in the initial part
of  the time  period.  This would  prevent the  adequate estimation of
this component  if a model with fixed  periodicity  was used.  In contrast
the method used here allows for slow variations in the amplitudes
of the oscillations in addition to variations in frequency.  A strong
stimulus response is noticeable in this  principal component (data set
\C\ was gathered in presence of visual stimulus). 

\subsubsection{Space-Frequency SVD}

\setcounter{figure}{24}
\begin{figure}
\centerline{\hbox{\psfig{figure=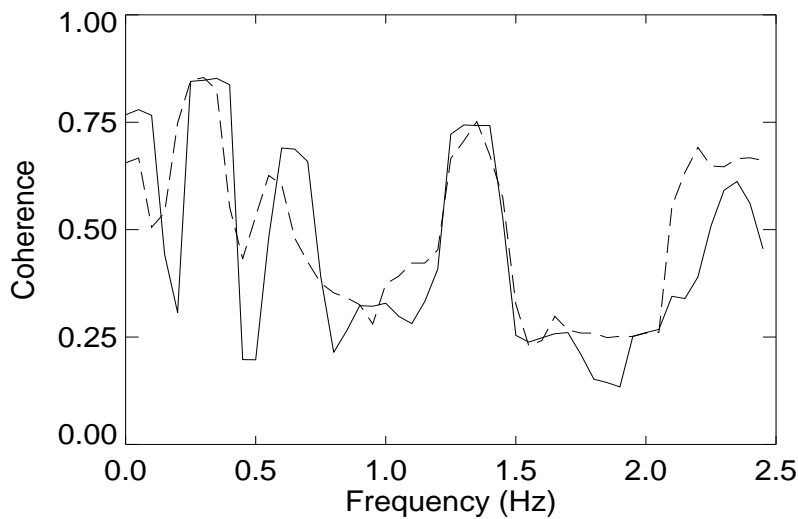,height=3in,width=4.5in}}}
\caption{
Coherence spectrum for a space-frequency SVD of fMRI data
from data sets \C\ and \D\ in presence of visual stimulus (solid curve) 
compared with absence of visual stimulus (dotted curve). 
}
\end{figure}

 In this  section, the results of a  space-frequency SVD are shown for
 data sets \C\ and \D.  Recall that data sets \C\ and \D\ were collected
with
 identical protocols,  except that in data set  \C\ a  controlled visual
 stimulus  is applied.  Fig.25  shows the  coherence spectra resulting
 from the space-frequency  SVD.   In this calculation,  the  DPSS used
 corresponded to  a full bandwidth  of $0.1Hz$. The  coherence spectra
 for the two data sets  are more or less the  same. The coherence near
 zero frequency is higher for  data set \C\   which contains the  visual
 stimulus. The stimulus response can be seen clearly in the amplitude
 of the leading spatial eigenmode of the space-frequency SVD for data
 set \C\ (Fig.26) at  close  to zero  frequency. At higher  frequencies,
 coherence arising  from  artifactual (respiratory) sources   causes a
 different pattern of spatial amplitudes. As opposed to the space-time
 SVD,   this  procedure segregates   the   stimulus response from  the
 oscillatory artifacts.

\setcounter{figure}{25}
\begin{figure}
\centerline{\hbox{\psfig{figure=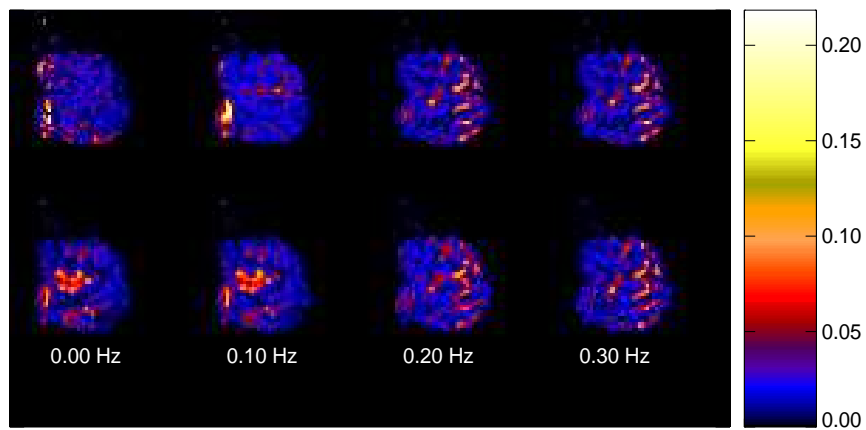,height=4in,width=4.5in}}}
\caption{Amplitudes of leading spatial eigenmodes corresponding to the
space-frequency SVD in Fig.25.  Center frequencies from $0$ Hz to
$0.3$ Hz in $0.1$ Hz increments.  Upper:  Without stimulus.  Lower:
With stimulus.
}
\end{figure}

\section{Discussion} 

We have tried to outline analysis protocols for data from the 
different modalities of brain imaging. It is useful recapitulate 
the essential features of these protocols in a unified manner,
and to indicate the domains of validity of the different techniques 
proposed.  

\noindent{\it Visualization of raw data: } \\
It is usually necessary to directly
visualize the raw data, both as a crude check on the quality of 
the experiment, and to direct further analysis. In this stage, 
one may look at individual time series from the images, or look 
at the data displayed dynamically as a movie. The relevant images
are often noisy, so that a noise reduction step is first necessary
even before the preliminary visualization. In cases where the 
visualization is limited by large shot noise, truncation of a 
space-time SVD with possibly some additional smoothing provides 
a simple noise reduction step for the visualization.

\noindent{\it Preliminary characterization: } \\
In the next stage, it is useful 
to obtain quantities that help parse out the content of the data, 
in particular to identify the various artifacts. Despite its 
limitations, a space-time SVD is useful at this stage to reduce 
the data to a few time series and corresponding eigenimages. 
Examination of the aggregate spectra of the PC time series, for 
example, reveals the extent of cardiac/respiratory content of 
fMRI/optical imaging data. In case of MEG, direct examination 
of the PC time series reveals the degree of cardiac contamination. 
Examination of the corresponding spatial images reveals the 
spatial locations of the artifacts. In case of fMRI data, where
the digitization rate may not be very high, studying the spectra
can reveal whether cardiac/respiratory artifacts still lead to 
possibly aliased frequency peaks in the power spectra. 

A further, more powerful characterization is obtained by the 
space-frequency SVD. For optical data and for rapidly sampled 
fMRI data, there is sufficient frequency resolution that at this 
stage the oscillatory artifacts segregate well. Studying the 
overall coherence spectrum reveals the degree to which the images are
dominated by the respective artifacts at the artifact frequencies, 
while the corresponding leading eigenimages show the spatial 
distribution of these artifacts more cleanly compared to the 
space-time SVD. Moreover, provided the stimulus response does not 
completely overlap the artifact frequencies, a characterization is
also obtained of the spatio-temporal distribution of the 
stimulus response. In case of fMRI, if the digitization rates 
are too slow (say less than 0.3Hz), there may not be any segregation
in the frequency domain of the various components of the image; 
this can be established at this stage by examining the eigenimages
of the space-frequency SVD. In this case, the techniques described
in this paper would be of limited use. 

\noindent{\it Artifact removal: }\\
Based on the preliminary inspection stage, one can proceed to 
remove the various artifacts to the extent possible. The techniques 
described in this paper are most relevant to artifacts that are 
sufficiently periodic, such as cardiac/respiratory artifacts in
optical/fMRI data, 60 cycle noise in optical data/MEG,
other frequency-localized noise such as building/fan vibrations
(optical imaging data). There are two basic ways of using the 
frequency segregation of the artifacts to remove them. One method is 
to directly model the waveforms of the oscillations using the 
frequency and amplitude modulated sinusoidal fit described in
the sections on optical/fMRI data. 

For fMRI data, if the digitization
rate is too low, then the techniques described here are not 
useful. However, for fMRI data, even with 
digitization rates of 1-2 seconds, 
it appears possible to use the frequency segregation of the 
physiological artifacts, using one of two methods. If auxiliary 
time series are available for cardiac/respiratory oscillations, 
one may construct the transfer function from these time series
to the data using the multitaper technique described above, 
perform a statistical test of significance such as the F-test, 
and remove the significantly fitted components. Alternatively, 
if no auxiliary data are available, the space-frequency SVD 
may be examined for presence of these artifacts, and if 
the artifacts can be identified with some frequency band then 
filtering techniques may be used. 

In each of the cases described above, the fundamental operation
is performed on an individual time series. This may be performed 
pixel by pixel in the image, or to reduce computational time, 
the steps may alternatively be performed on the leading PC time 
series and the artifacts thus reconstructed may be then subtracted
from the raw data. 

\noindent{\it Stimulus response characterization:} \\
This is may be the most delicate step, since the goal of the 
experiment is usually to find the stimulus response which is 
not known {\it a priori}. If the stimulus is presented 
periodically and repeatedly, the transfer function may be 
computed in the frequency domain using the techniques described
before. Often, some strong assumption is made about the 
stimulus response ({\it e.g.} the image intensity will rise
during the stimulus), and methods of signal detection theory
or statistical hypothesis testing are applied to extract the response
based on the assumed signal model. 

One general assumption about the stimulus response to a single 
trial might be that it lives in a particular region of frequency
space. In the case of fMRI data, where the response is often a 
prolonged increase in signal intensity, this would correspond 
to the signal having relatively low frequencies. In this case, 
the space-frequency SVD described in the paper is of utility 
in describing the stimulus response, as illustrated in the section
on fMRI data. Similarly, in optical imaging, if the stimulus 
was modulated with a particular temporal frequency, the same 
idea would be applicable.

It is to be emphasized that it is not  possible to develop "black box"
like   techniques which    are a   panacea  to all    problems of data
analysis. It is neither possible  nor desirable to entirely  eliminate
the human  component  in the process. However,  relevant computational
and analytical tools can be  a  powerful aid to   making sense of  the
data, and are used most effectively in a closed  loop system where the
results of analysis influence experimentation.

\section*{Acknowledgements}
The authors gratefully acknowledge the contribution of published as
well as unpublished data from several colleagues, including
K. R. Delaney, A. Gelperin, X. Hu, D. Kleinfeld, R.  Llinas, 
S. Ogawa, U. Ribary, and K. Ugurbil.  We are indepted to D.J. 
Thomson for help with multitaper
spectral techniques. We thank K. Svoboda, D.Kleinfeld and
L.Cohen for comments on the manuscript. We would also like to thank two 
anonymous referees for extensive comments that led to substantial 
improvements in the manuscript. This research was supported 
by Bell Laboratories, Lucent Technologies
(Internal funding).

\bibliographystyle{apalike}
\bibliography{bpbib}

\end{document}